\documentclass[prm,showpacs,twocolumn,longbibliography]{revtex4-1}
\usepackage[utf8]{inputenc}
\usepackage{graphicx}
\usepackage{textcomp}
\usepackage{txfonts}
\usepackage{url}
\urldef{\SMonline}\url{http://link.aps.org/supplemental/10.1103/PhysRevMaterials.7.104402}
\usepackage{color}
\usepackage{gensymb} 
\usepackage{tikz}
\usetikzlibrary{positioning} 
\usetikzlibrary{calc} 
\usetikzlibrary{shapes}
\usetikzlibrary{arrows}

\renewcommand{\vec}[1]{{\bf #1}}

\newcommand{\jena}{Institute of Condensed Matter Theory and Optics, Friedrich Schiller University Jena, Max-Wien-Platz 1, 07743 Jena, Germany}

\begin{document}

\title{Optical signatures of defects in BiFeO$_3$}

\author{Sabine K\"{o}rbel}
\email{skoerbel@uni-muenster.de}
\affiliation\jena

\begin{abstract}
%
%
	Optical absorption in rhombohedral BiFeO$_3$ starts at photon energies below the photoemission band gap of $\approx$3~eV calculated from first principles.
	A shoulder at the absorption onset has so far been attributed to low-lying electronic transitions or to oxygen vacancies.
	In this work optical spectra are calculated {\textit{ab initio}} to determine the nature of the optical transitions near the absorption onset of pristine BiFeO$_3$, 
	the effect of electron-hole interaction,
	and 
	the spectroscopic signatures of typical defects, i.e. doping (excess electrons or holes), intrinsic defects (oxygen and bismuth vacancies), and low-energy structural defects (ferroelectric domain walls). 
%
\end{abstract}

\maketitle

\section{Introduction}

BiFeO$_3$ is a well-studied ferroelectric material with a very stable and large ferroelectric polarization
close to 100~$\mu$C/cm$^2$~\cite{lebeugle:2007:very}, a high ferroelectric Curie temperature above 1100~K~\cite{smolenski:1982:ferroelectromagnets},
and a relatively small direct optical band gap, compared to other ferroelectric oxides,
of $\approx$2.7--3.0~eV \cite{basu:2008:photoconductivity,hauser:2008:characterization,ihlefeld:2008:optical,kumar:2008:linear,zelezny:2010:optical,sando:2018:revisiting,moubah:2012:photoluminescence}.

BiFeO$_3$ exhibits a bulk photovoltaic effect due to its non-centrosymmetric $R3c$ crystal structure 
\cite{fridkin:2001:bulk,choi:2009:switchable,ji:2011:evidence,bhatnagar:2013:role}.

Because of its advantageous ferroelectric and optical properties, BiFeO$_3$ may be considered a starting point for combining ferroelectric and optoelectronic functionality,
even though its band gap is still too large to efficiently harvest visible light.


In spite of a large number of publications on the subject, the optical properties of BiFeO$_3$ and the role of defects are still not completely understood.
A fundamental or photoemission band gap between 3.0 and 3.4~eV \cite{goffinet:2009:hybrid,stroppa:2010:hybrid} was found using accurate first-principles calculation methods, namely hybrid functionals or the $GW$ approximation.
Optical absorption and luminescence spectra show a peak at about 3~eV, in agreement with the calculation results, but there is a shoulder at about 2.5~eV \cite{basu:2008:photoconductivity,hauser:2008:characterization,pisarev:2009:charge,choi:2011:optical, 
hauser:2008:characterization,moubah:2012:photoluminescence}, which has 
been suggested to arise from the electronic structure of BiFeO$_3$ itself (possibly from excitons) \cite{basu:2008:photoconductivity,pisarev:2009:charge}, 
or from oxygen vacancies \cite{hauser:2008:characterization,himcinschi:2010:substrate,moubah:2012:photoluminescence}. 
Note that the assumed origin of the shoulder (intrinsic versus defects) in previous works does not coincide with the approach taken (experiment versus theory).
With first-principles calculations of optical absorption spectra in the independent particle approximation (excluding electron-hole interaction) a shoulder at the absorption onset was found \cite{lima:2020:optical}, 
indicating that the shoulder originates in the electronic structure of pristine BiFeO$_3$.
In a different study \cite{radmilovic:2020:combined} using similar computational methods the shoulder was not visible.

First principles studies of BiFeO$_3$ with oxygen vacancies revealed electronic states below the conduction band minimum \cite{ederer:2005:influence,ju:2009:first,radmilovic:2020:combined} 
that might explain the shoulder.
To be visible in absorption, oxygen vacancies would need to be present in concentrations near 10$^{20}$/cm$^3$, which seems high but possible. 
If the shoulder arises from $V_{\mathrm{O}}$, its magnitude should depend on the concentration of $V_{\mathrm{O}}$ and lose intensity upon annealing in oxygen.
Another candidate for gap states could be the Bi vacancy. 
The Bi vacancy was found to introduce holes on Fe, forming Fe$^{4+}$, in the vicinity of ferroelectric domain walls \cite{rojac:2017:domain}.
\newline
\indent Not only atomic, but also electronic defects such as excess electrons or electron holes may create extra peaks in optical absorption spectra.
Both structurally and electronically, BiFeO$_3$ is similar to hematite, Fe$_2$O$_3$. 
Both materials contain Fe atoms in the formal charge state Fe$^{3+}$ inside oxygen octahedra; 
for both materials the electronic structure near the band edges is mainly composed of O$-p$ and Fe$-d$ states. 
In the case of hematite, there is evidence indicating that excess electrons, introduced via doping, tend to localize and form polarons with 
an
electronic state inside the band gap 
 well-separated from the conduction-band bottom \cite{lohaus:2018:limitation,peng:2012:semiconducting}.
A similar behavior may also be expected in BiFeO$_3$ \cite{koerbel:2018:electron}.
Previous theoretical first-principles studies found 
that excess electrons in BiFeO$_3$  become self-trapped into small polarons \cite{koerbel:2018:electron}, whereas
holes form large polarons only \cite{koerbel:2018:electron,geneste:2019:polarons}.
\newline
\indent Besides electronic and atomic defects, also structural defects, such as ferroelectric domain walls, are often present and should be considered when discussing material properties.
In fact, ferroelectric domain walls may even be desirable, as they can bring additional functionality, such as enhanced electrical conductivity compared to the domain interior \cite{seidel:2009:conduction,guyonnet:2011:conduction,farokhipoor:2011:conduction,bencan:2020:domain}.
%
The ferroelectric domain walls may act as traps for atomic \cite{gaponenko:2015:towards,rojac:2017:domain} and electronic defects \cite{koerbel:2018:electron}.
\newline
\indent The purpose of the present work is to determine the origin of the shoulder at the optical absorption onset and the effect of electron-hole interaction on the optical absorption spectrum of BiFeO$_3$, and to identify the signatures of defects.
\newline
\indent To this end the optical absorption spectrum of BiFeO$_3$ was modeled {\textit{ab initio}} including electron-hole effects by means of density-functional theory (DFT) with $GW$ corrections of the electronic energies 
and by subsequently solving the Bethe-Salpeter equation (BSE). 
The resulting optical spectra are compared to published ellipsometry data, the electronic transitions responsible for the peaks near the absorption onset are identified, and the excitonic binding energy is determined.
A computationally less demanding approach, the independent particle approximation on DFT+$U$ level, is validated for defect-free BiFeO$_3$ and subsequently used to determine the optical spectra of BiFeO$_3$ with defects,
namely, electron and hole polarons, oxygen and Bi vacancies, and ferroelectric domain walls, for which the $GW$+BSE approach would be computationally too demanding.

\section{\label{sec:methods}Methods}

\subsection{Ground-state properties}
The calculations were performed with the program {\sc{vasp}} \cite{kresse:1996:efficiency},
using the projector-augmented wave method and pseudopotentials with 5 (Bi), 16 (Fe), and 6 (O) valence electrons, respectively (Bi, Fe-sv, O, PSP0) (78 valence electrons per unit cell in total).
The local spin-density approximation (LSDA) to DFT was used, 
and the band gap was corrected with a Hubbard $U$ of 5.3~eV applied to the Fe-$d$ states using Dudarev's scheme \cite{dudarev:1998:electron}.
This $U$ value was taken from the Materials Project \cite{jain:2013:materials}; 
it was optimized with respect to oxide formation energies, but also yields band gaps and ferroelectric polarization close to experiments \cite{koerbel:2020:photovoltage}.
With $U$=5.3~eV, a ferroelectric polarization of $\approx 94\,\mu$C/cm$^2$ \cite{koerbel:2020:photovoltage} is obtained (Expt.: $\approx 100\,\mu$C/cm$^2$ \cite{lebeugle:2007:very}).
The reciprocal space was sampled with a $k$-point mesh of $8\times8\times8$ points in the first Brillouin zone of the 10-atom unit cell of the $R3c$ phase. 
Plane-wave basis functions with energies up to 520~eV were used.
Both the atomic positions and the cell parameters were optimized until 
the total energy differences between consecutive iteration steps 
fell below 0.01~meV for the optimization of the electronic density and below 0.1~meV for the optimization of the atomic structure.

\subsubsection{BiFeO$_3$ with defects}
BiFeO$_3$ with defects was modeled using supercells with 120 atoms employing periodic boundary conditions.
These supercells were identical to those used in Ref. \cite{koerbel:2018:electron}.
More precisely, for the $109\degree$ wall, the supercell is spanned by ($\vec{a}_{\mathrm{pc}}+\vec{b}_{\mathrm{pc}}$), 
($\vec{a}_{\mathrm{pc}}-\vec{b}_{\mathrm{pc}}$), and 12$\vec{c}_{\mathrm{pc}}$, where $\vec{a}_{\mathrm{pc}}$, $\vec{b}_{\mathrm{pc}}$, and $\vec{c}_{\mathrm{pc}}$ span the pseudocubic five-atom cell. 
In the case of the $71\degree$ and the $180\degree$ walls, the supercell is spanned by 2$\vec{a}_{\mathrm{pc}}$, ($\vec{b}_{\mathrm{pc}}-\vec{c}_{\mathrm{pc}}$), and 6($\vec{b}_{\mathrm{pc}}+\vec{c}_{\mathrm{pc}}$).
The subject of this paper is a two-dimensional array of defects accumulated in a plane, not an isolated defect. 
The dimensions of the supercell are suitable to model exactly that.
%
In the case of charged defects, electrons were added or removed and a uniform compensating background charge density was added. 
Spin-orbit coupling was neglected.
\paragraph*{Atomic defects.}
Several initial positions were considered and the structures optimized,
then the lowest-energy configuration was selected for further investigation.
For modeling the hole at the Bi vacancy, an additional Hubbard $U$ of 8~eV \cite{sadigh:2015:variational,erhart:2014:efficacy} was applied to the $p$ orbitals of O 
 to remove self-interaction errors in the hole state.
\paragraph*{Electronic defects.}
Throughout this paper, similar to Ref.~\cite{franchini:2021:polarons},  small and large polarons are distinguished by the spatial extension of the electronic wave function compared to the interatomic distances 
(small polaron: localized within maximally a few interatomic distances; large polaron: delocalized over $\geq$ tens of interatomic distances).
The abbreviations SEP and LHP are used for the small electron polaron and for the large hole polaron, respectively.

\paragraph*{Ferroelectric domain walls.}
The atomic structures and formation energies of low-energy ferroelectric domain walls in BiFeO$_3$ 
are well-known \cite{seidel:2009:conduction,dieguez:2013:domain,ren:2013:ferroelectric,wang:2013:bifeo3,chen:2017:polar}.
Three types of domain walls can form with different angles between the polarization directions in adjacent domains:
71\textdegree, 109\textdegree, and 180\textdegree, all of which are experimentally observed \cite{seidel:2009:conduction,rojac:2017:domain,wang:2013:bifeo3}. 
Further details regarding the properties of different ferroelectric domain walls in BiFeO$_3$ can be found in Ref.~\cite{dieguez:2013:domain}.
Only charge-neutral and ``mechanically compatible'' \cite{fousek:1969:orientation,dieguez:2013:domain} 
ferroelectric domain walls in lattice planes with low Miller indices as presumably the most abundant types of domain walls are considered here.
The atomistic models of ferroelectric domain walls were 
obtained by creating a supercell of ferroelectric BiFeO$_3$ in the $R3c$ phase and changing the polarization direction in 
one-half of the supercell to one of the other equivalent directions 
by applying an appropriate symmetry operation to the atom positions in half of the  
supercell and subsequently optimizing the atomic coordinates and lattice parameters. 
In this way, atomic coordinates and formation energies of domain walls are obtained that are in close agreement 
with both experimental and other theoretical work \cite{wang:2013:bifeo3,dieguez:2013:domain,chen:2017:polar,ren:2013:ferroelectric}.
For the supercell calculations with 120 atoms, a $k$-point mesh was used that corresponds to $5\times 5\times 5$ $k$-points for the ten-atom unit cell. 
More details, including convergence tests and comparison with the literature, are given in the Supplemental Material of Ref.~\cite{koerbel:2020:photovoltage}.
%
%
Atomic and electronic structure of the domain-wall systems investigated here 
have been published elsewhere \cite{dieguez:2013:domain,wang:2013:bifeo3,koerbel:2018:electron}.
%
%
The antiferromagnetic $G$-type spin configuration of the bulk was maintained in the systems with domain walls.

\subsection{Optical properties}
\subsubsection{Pristine BiFeO$_3$}
Optical properties of pristine, perfect BiFeO$_3$ were determined using many-body perturbation theory.
First, a one-shot $GW$ ($G_0W_0$) calculation was performed for the lowest $\geq$110 bands 
to obtain electronic energy eigenvalues 
that include electron-electron interaction beyond the 
mean-field approach of DFT+$U$, starting from LSDA+$U$ ($U$=6~eV) with a $k$-mesh of $6\times 6\times 6$ points, 
using 150 frequencies, an energy cutoff of 300~eV for the dielectric permittivity, and $\approx 2000$ bands in total.
Pseudopotentials were chosen for which core electrons are well separated in energy from valence electrons (see Fig.~17 in the Supplemental Material \cite{Note1}).
Two different sets of pseudopotentials were employed (for details, see Table~\ref{tab:calc_par_BSE} and the Supplemental Material \footnote{See Supplemental Material at \SMonline\ for a validation of the computational methods, fits, and additional computational details \label{footnote:SM}}) and the results compared. 
Then the dielectric permittivity including electron-hole effects was calculated using the BSE.
The BSE was solved for 16 valence and 16 conduction bands, employing the Tamm-Dancoff approximation. 
The calculation parameters adopted for obtaining optical properties of pristine BiFeO$_3$ are compiled in Table~\ref{tab:calc_par_BSE}.
Similar to Refs.~\cite{bokdam:2016:role,varrassi:2021:optical}, a computationally lighter approach using a model dielectric function in the BSE was used to obtain optical spectra for denser $k$-point meshes, 
which were then extrapolated to an infinitely dense $k$-point mesh by a fitted power law. 
The model for the dielectric function reads $\varepsilon_{\vec{G},\vec{G}}^{-1}(\vec{q})=1-(1-\varepsilon_{\infty}^{-1})\exp(-|\vec{q}+\vec{G}|^2/4\lambda^2)$ \cite{bokdam:2016:role}.
The dielectric constant $\varepsilon_{\infty}\approx$7.3 was calculated with LSDA+$U$, the exponential decay parameter $\lambda$=1.4--1.5~\AA$^{-1}$ was obtained by fitting (see Sec. H and Table~I in the Supplemental Material \cite{Note1}).
Before solving the model BSE, a scissor of $\approx$0.8--1.4~eV was employed on top of the Hubbard $U$ \cite{Note1} (mBSE@LSDA+$U$+scissor). 
The scissor was chosen such that the $G_0W_0$ gap is reproduced.
The gaps obtained at the (BSE@)$G_0W_0$@LSDA+$U$ level were {\textit{a posteriori}} corrected for spin-orbit coupling ($-$0.1~eV as obtained with LSDA+$U$)
and extrapolated to an infinite energy cutoff for the dielectric matrix \cite{qiu:2013:optical} and an infinitely dense $k$-point mesh (BSE only).
A detailed validation of the computational methodology can be found in the Supplemental Material \cite{Note1}. 
\begin{table}
	\caption{\label{tab:calc_par_BSE}Calculation parameters of the $GW$ and BSE calculations for pristine BiFeO$_3$. 
	SOC: Spin-orbit coupling, LFE: Local-field effects.}
	\footnotesize{
	\begin{tabular}{l  l}\hline\hline
	Theory level                      &   BSE@$G_0W_0$@LSDA+$U$ \\
	Pseudopotentials                  &  PSP1, PSP2 \\
		$U$                       &   6~eV     \\
		Pseudopotential 1 (PSP1)  &  Bi-d, Fe, O\\
		Pseudopotential 2 (PSP2)  &  Bi-sv-GW, Fe-sv-GW, O-GW\\
       $E_{\mathrm{cut}}$                 & 300~eV \\
	Number of bands                   & $\geq$2000  \\
Number of bands corrected with $G_0W_0$   & $\geq$100  \\
Number of bands in the BSE                & 16 VB, 16 CB  \\
 	$N_{\omega}$                      &  150  \\
$E_{\mathrm{cut}}(\varepsilon)$           & 300~eV \\
    $k$-points                            & $6 \times 6\times 6$ \\
		SOC                       &   included {\textit{a posteriori}}  \\
		LFE                       &   included \\ \hline\hline
	\end{tabular}
}
\end{table}
%
\subsubsection{BiFeO$_3$ with defects}
BiFeO3 with defects was modeled using the computational setup summarized in Table\ref{tab:calc_par_defects}.
Since modeling the systems with defects is computationally expensive already at the DFT level, 
many-body perturbation theory was not employed to calculate their optical properties.
Instead the independent-particle approximation (IPA) on the LSDA+$U$ level was used to calculate 
the frequency-dependent imaginary part of the high-frequency relative dielectric permittivity, 
$\varepsilon_2(\omega)$, thereby neglecting excitonic effects and local-field effects. 
The justification for this approach lies in its being computationally light-weight enough to be applied to supercells.
In the energy region within 2~eV above the absorption onset, the local-field effects mainly reduce $\varepsilon_2$ by $\approx 10$\% without modifying the shape, see Fig.~15 in the Supplemental Material \cite{Note1}. 
In this case, $\varepsilon_2$ is given by
\begin{eqnarray}
  \varepsilon_2^{\alpha\beta}(\omega)=&\frac{4\pi^2e^2}{\Omega}\lim\limits_{\vec{q}\to\vec{0}}\frac{1}{q^2}\sum\limits_{c,v,\vec{k}} 2 w_{\vec{k}}\delta(\epsilon_{c\vec{k}}-\epsilon_{v\vec{k}}-\omega)\nonumber\\
                              &\cdot \langle u_{c\vec{k}+\vec{e_{\alpha}}q}|u_{v\vec{k}}\rangle\langle u_{c\vec{k}+\vec{e_{\beta}}q}| u_{v\vec{k}}\rangle^*,
         \label{eq:eps2}
\end{eqnarray}
where $\alpha$ and $\beta$ are Cartesian directions, $v$ and $c$ are valence- and conduction-band indices at $k$-point $\vec{k}$,
the $\epsilon_{n\vec{k}}$ are energy eigenvalues, $u$ is the lattice-periodic part of the Bloch function, $\vec{e}$ is a Cartesian unit vector,
$\Omega$ is the unit-cell volume, and $w_{\vec{k}}$ is the $k$-point weight.
The real part of the dielectric permittivity $\varepsilon_1$, needed to calculate the absorption coefficient, 
was obtained from a Kramers-Kronig transformation \cite{gajdos:2006:linear}.
\noindent 
Where not specified otherwise,
a broadening of 0.1~eV
was applied to the optical spectra to mimick broadening effects present in experimental spectra. 
The absorption coefficient $\alpha$ is given by 
 $ \alpha=\frac{4\pi \nu\kappa}{c}$,
\noindent where $\nu$ is the frequency of the incident light, $c$ is the speed of light, 
and $\kappa=\sqrt{(|\varepsilon| -\varepsilon_1)/2}$ 
is the imaginary part of the complex index of refraction \cite{jackson:1999:classical}.
To obtain a direction-averaged dielectric permittivity and absorption coefficient,
the eigenvalues of the dielectric permittivity matrix, $\varepsilon_2$, are averaged.
\newline
\indent To model photoluminescence (PL), spontaneous emission from electron-hole pairs is assumed that are in thermal quasi-equilibrium in the excited state, but not in thermal equilibrium with the photon field. 
Only the zero-phonon line is considered here.
The intensity of the photoluminescence as a function of frequency is here approximated by \cite{einstein:1917:zur,cardona:2005:fundamentals,hannewald:2000:theory}
\begin{eqnarray}
	I_{\mathrm{PL}}(\omega,T)\sim&\omega^3\lim\limits_{\vec{q}\to\vec{0}}\sum\limits_{c,v,\vec{k}} w_{\vec{k}} f_{c\vec{k}}(T,\mu_e)\left(1-f_{v\vec{k}}(T,\mu_h)\right)\nonumber\\
			      &\delta(\epsilon_{c\vec{k}}-\epsilon_{v\vec{k}}-\omega) \cdot \langle u_{c\vec{k}+\vec{q}}|u_{v\vec{k}}\rangle\langle u_{c\vec{k}+\vec{q}}| u_{v\vec{k}}\rangle^*,
         \label{eq:PL}
\end{eqnarray}
where $f_{c\vec{k}}$ and $f_{v\vec{k}}$ are the Fermi distributions in the excited state,
$f_{v/c\vec{k}}=1/(e^{(\varepsilon_{v/c\vec{k}}-\mu_{h/e})/k_B T}+1)$;
$\mu_h$, $\mu_e$ are the quasi-Fermi levels of holes and electrons; and $T$ is the temperature. 
$\mu_h$ and $\mu_e$ were determined based on the electronic density of states and chosen such that there is one excess electron and one hole in the 120-atom supercell.
Equation (\ref{eq:PL}) was implemented in a home-made code \cite{cofimaker:github}.
To take into account defect concentrations below those accessible in supercell calculations, 
here the photoluminescence signal for a defect concentration $c_{\mathrm{defect}}$ is approximated by 
$\left[N_{\mathrm{deg}}^{\mathrm{bulk}} I_{\mathrm{PL}}^{\mathrm{bulk}}(\omega,T)+c_{\mathrm{defect}} N_{\mathrm{deg}}^{\mathrm{defect}}  I_{\mathrm{PL}}^{\mathrm{defects}}(\omega,T)\right] d(\omega,T),$
where $d(\omega,T)$ is the thermal distribution function for the electron-hole pairs with excitation energy $\hbar\omega$, and $N_{\mathrm{deg}}^{\mathrm{bulk/defect}}$ is the degeneracy of the bulk/defect level involved in the transition. Here $N_{\mathrm{deg}}^{\mathrm{defect}}\approx N_{\mathrm{deg}}^{\mathrm{bulk}}$ is assumed.
$I_{\mathrm{PL}}^{\mathrm{bulk}}$ and $I_{\mathrm{PL}}^{\mathrm{defect}}$ are PL intensities calculated for the same number of excess electrons or holes and the same supercell size.
For independent electrons and holes, $d(\omega,T)$ is the Fermi-Dirac distribution $1/(e^{\hbar\omega/k_B T}+1)$, 
for excitons it is the Bose-Einstein distribution $1/(e^{\hbar\omega/k_B T}-1)$ \cite{torun:2018:interlayer,cannuccia:2019:theory}.
Since the calculated exciton binding energy $E_B(X)\gg k_BT$ at room temperature, here the Bose-Einstein distribution is used.
\newline
\indent Both recombination from an electron polaron with an instantaneous hole and from a small exciton polaron were considered; the latter was modeled using excitonic $Delta$ self-consistent field ($\Delta$SCF), i.e. 
by enforcing simultaneously a hole in the valence bands and an electron in the conduction bands in a DFT calculation during geometry optimization.
References and details regarding the implementation are given in Ref.~[\onlinecite{koerbel:2020:photovoltage}].
The pseudopotentials used were the same as in the ground state calculations (PSP0).
\begin{table}
	\caption{\label{tab:calc_par_defects}Calculation parameters for BiFeO$_3$ with defects. 
	IPA: Independent particle approximation, SOC: Spin-orbit coupling, LFE: Local-field effects.
	``$\hat{=}5 \times 5\times 5$'' means the $k$-point spacing for the supercell is as close 
	as possible to a $5 \times 5\times 5$ mesh for the 10-atom unit cell.}
	\begin{tabular}{l  l}\hline\hline
	Theory level        &   LSDA+$U$-IPA \\
	Pseudopotentials     & PSP0 \\
		$U$         &   5.3~eV     \\
		$k$-points              & $\hat{=} 5 \times 5\times 5$ \\
		SOC         &   neglected  \\
		LFE         &   neglected  \\\hline\hline
	\end{tabular}
\end{table}
\section{\label{sec:results}Results and Discussion}
%
\subsection{Optical spectra of pristine BiFeO$_3$}
\begin{figure}
  \includegraphics[width=0.49\textwidth]{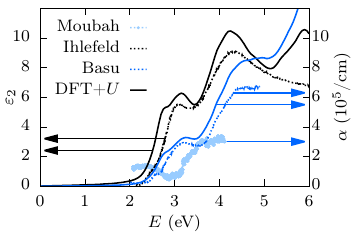}
  \caption{\label{fig:eps} Imaginary part of the dielectric permittivity 
    $\varepsilon_2$ (left axis) and absorption coefficient $\alpha$ (right axis) 
    from DFT+$U$ ($U$=5.3~eV, solid lines) and experiment (dotted lines). 
    Experimental spectra are taken from 
	Refs.~[\onlinecite{ihlefeld:2008:optical}] ($\varepsilon_2$),
	and [\onlinecite{basu:2008:photoconductivity}], and [\onlinecite{moubah:2012:photoluminescence}] ($\alpha$).
	The data of Ref.~[\onlinecite{moubah:2012:photoluminescence}] for $\alpha$ (thick blue dotted line) are in in arbitrary units.
    }
\end{figure}

The bandgaps of BiFeO$_3$ calculated with different levels of theory as reported in the literature and from this work are compiled in Table~\ref{tab:BFO_gaps_theory}.
\begin{table}
	\caption{\label{tab:BFO_gaps_theory} Band gaps of BiFeO$_3$ obtained with different levels of theory. The band gaps from $G_0W_0$ and BSE@$G_0W_0$ in this work strongly depend on the pseudopotentials used (see Fig.~18 in the Supplemental Material \cite{Note1}).}
	{\footnotesize{
	\begin{tabular}{lc}\hline\hline
		Method                                                               & Gap (eV)              \\ \hline
		$G_0W_0$ (this work)                                                 & 3.1--3.6 (fund.); 3.1--3.7 (direct)     \\
		BSE@$G_0W_0$  (this work)                                            & 2.9--3.5       \\
		$E_B(X)$ (this work)                                                 & 0.2                       \\
		Hybrid DFT (HSE, Ref.~[\onlinecite{stroppa:2010:hybrid}])              & 3.4                    \\
		Hybrid DFT (B1-WC, Ref.~[\onlinecite{goffinet:2009:hybrid}])           & 3.0                    \\
		$G_0W_0$@HSE (Ref.~[\onlinecite{stroppa:2010:hybrid}])                 & 3.8                    \\
		$GW$ with vertex corrections (Ref.~[\onlinecite{stroppa:2010:hybrid}]) & 3.3                    \\
		Experiment (optical) (Refs.~[\onlinecite{basu:2008:photoconductivity,hauser:2008:characterization,ihlefeld:2008:optical,kumar:2008:linear,zelezny:2010:optical,sando:2018:revisiting,moubah:2012:photoluminescence}])                                           & 2.7--3.0\\\hline\hline
	\end{tabular}
	}}
\end{table}
First, the 
calculated optical spectrum of the pristine bulk crystal is compared with the experimentally measured one.
Figure~\ref{fig:eps} shows the imaginary part of the relative high-frequency dielectric permittivity (left axis)
and the absorption coefficient (right axis).
The calculated spectrum was obtained on LSDA+$U$ level ($U$=5.3~eV) in the IPA and averaged over the cartesian directions.
This approach is used below for the optical spectra of BiFeO$_3$ with defects.
%
\noindent
The calculated spectra 
on the independent-particle level are reasonably close to the experimentally measured ones of Refs. [\onlinecite{ihlefeld:2008:optical,basu:2008:photoconductivity}].
The underestimation of the quasiparticle gap and the missing excitonic binding energy and spin-orbit coupling largely cancel each other, 
except for a moderate redshift of $\approx 0.2$~eV.
The spectrum from Ref.~\cite{moubah:2012:photoluminescence} appears blueshifted by about 0.5~eV. %
It should be noted that a larger Hubbard $U$ would increase the optical gap in the independent-particle approximation and hence move it closer to experiment \cite{lima:2020:optical,ju:2009:electronic},
whereas including spin-orbit coupling would decrease the gap again \cite{lima:2020:optical}.
Considering that a larger $U$ value might overestimate the localization of electrons, a moderate $U$ of 5.3~eV, 
which yields simultaneously formation energies, band gaps, and ferroelectric polarization close to experiment (see Supplemental Material of Ref.~[\onlinecite{koerbel:2020:photovoltage}]), 
seems a safer choice for modeling polarons.

\begin{figure}
  \includegraphics[width=0.49\textwidth]{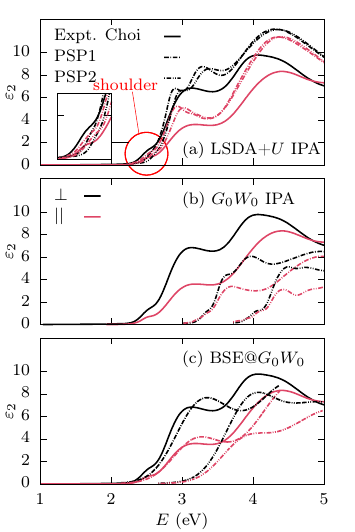}
  \caption{\label{fig:eps_direction_resolved} 
    Direction-resolved $\varepsilon_2$ from experiment \cite{choi:2011:optical}, 
	the independent particle approximation (IPA) on LSDA+$U$ level with $U$=6~eV (broadened dynamically by 0.01--1~eV), 
	the independent particle approximation on $G_0W_0$ level  (broadened dynamically by 0.001--1~eV),
	and the BSE (broadened by 0.4~eV) for light polarization perpendicular (black) and parallel (red) to the hexagonal $c$ axis (the rhombohedral [111] direction). 
	PSP1: Bi-d Fe O pseudopotential, PSP2: Bi-sv-GW Fe-GW O-GW pseudopotential.
  }
\end{figure}
Figure~\ref{fig:eps_direction_resolved} shows $\varepsilon_2$ for the ordinary ($\perp$ to the $c$ axis of the hexagonal unit cell) and extraordinary ($||\,c_{\mathrm{hex}}$) axes of BiFeO$_3$ 
from ellipsometry \cite{choi:2011:optical}, the IPA on the LSDA+$U$ level ($U$=6~eV), the IPA on the $G_0W_0$ level, and from the BSE@$G_0W_0$.
The shoulder at the absorption onset is less pronounced in the calculated spectra on the LSDA+$U$ level than in experiment, 
 on the $G_0W_0$-IPA level a shoulder is clearly visible. 
The magnitude of the shoulder depends on the broadening, the Hubbard $U$, 
and the pseudopotential
(see Figs.~1, 6, 10, and 11 in the Supplemental Material \cite{Note1}).
The BSE spectra are possibly not sufficiently converged with respect to $k$-points in order to resolve a shoulder, 
but the model BSE with more $k$-points yields a similar feature, see Fig. 12 in the Supplemental Material \cite{Note1}.

Figure~\ref{fig:BSE_states} shows the transition charge densities $\varrho_{e/h}^{(\lambda)}$ weighted with the dipole matrix element $\langle v\vec{k}|\vec{r}|c\vec{k}\rangle$ of the hole and electron states forming the lowest optical transition ($\lambda=1$) in the BSE,
\begin{equation}\label{eq:transition_density}
 \varrho_{e}^{(\lambda)}\sim\sum_{vc\vec{k}}|A_{vc\vec{k}}^{(\lambda)}|^2 |\langle v\vec{k}|\vec{r}|c\vec{k}\rangle|^2 |\psi_{c\vec{k}}|^2,
\end{equation}
and analogously for the hole,
where $\vec{A}^{(\lambda)}$ is the coupling coefficient for the BSE eigenvalue $E^{(\lambda)}$, so 
$\sum_{v'c'\vec{k'}}\mathcal{H}^{\mathrm{BSE}}_{vc\vec{k},v'c'\vec{k'}}A^{(\lambda)}_{v'c'\vec{k'}}=E^{(\lambda)}A^{(\lambda)}_{vc\vec{k}}$ \cite{rohlfing:2000:electron,onida:2002:electronic} 
and the excitonic wave function $\Psi_X^{(\lambda)}(\vec{r}_e,\vec{r}_h)=\sum_{cv\vec{k}}A^{(\lambda)}_{cv\vec{k}}\psi_{v\vec{k}}^*(\vec{r}_h) \psi_{c\vec{k}}(\vec{r}_e)$ \cite{rohlfing:2000:electron},
 where $\vec{r}_e$ and $\vec{r}_h$ are the positions of electrons and holes. 
 The indices $v$, $c$, and $\vec{k}$ denote valence bands, conduction bands, and $k$-points in the first Brillouin zone.
 The transition densities were calculated using PSP0.

\begin{figure}
  \includegraphics[width=0.25\textwidth]{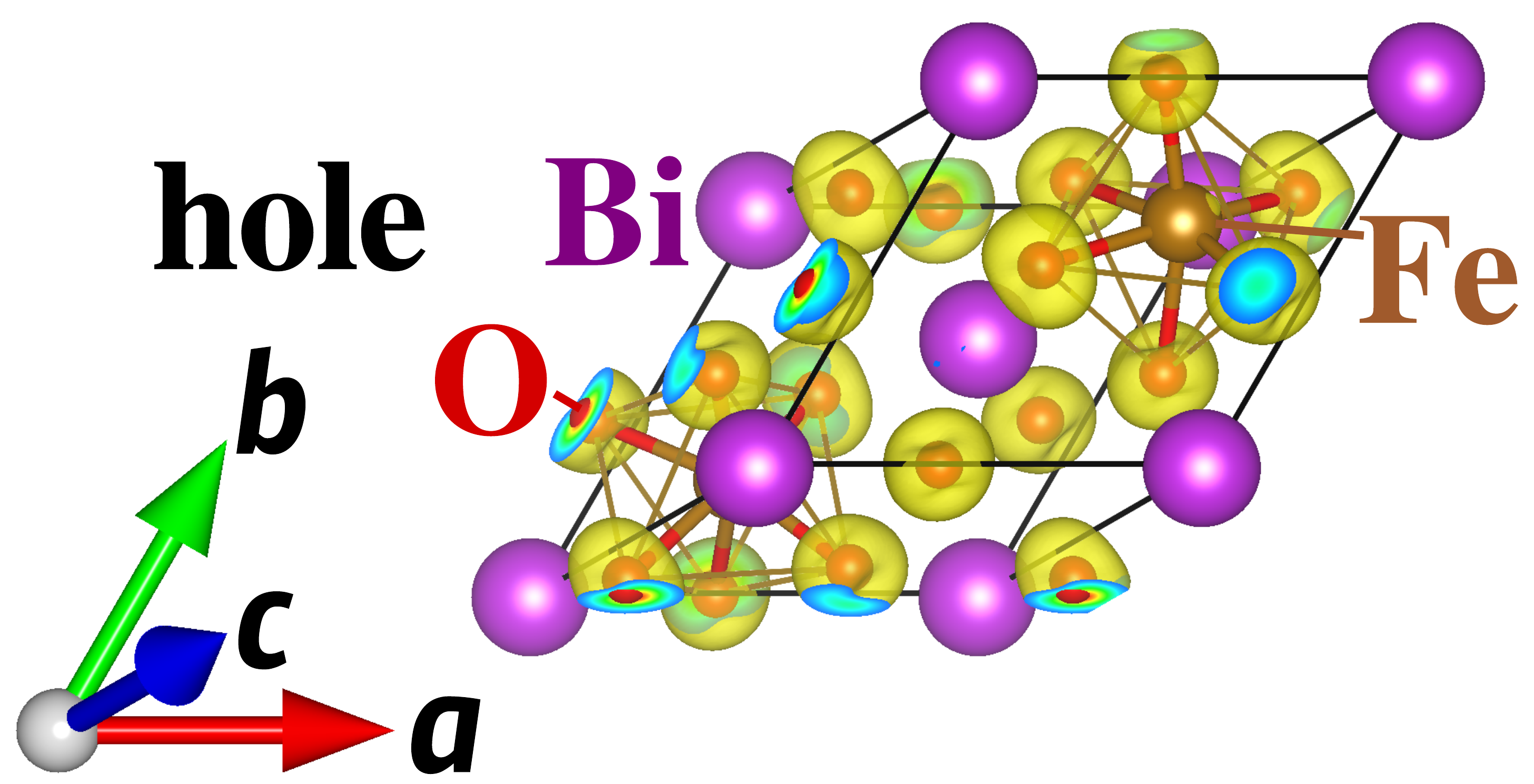}
  \includegraphics[width=0.2\textwidth]{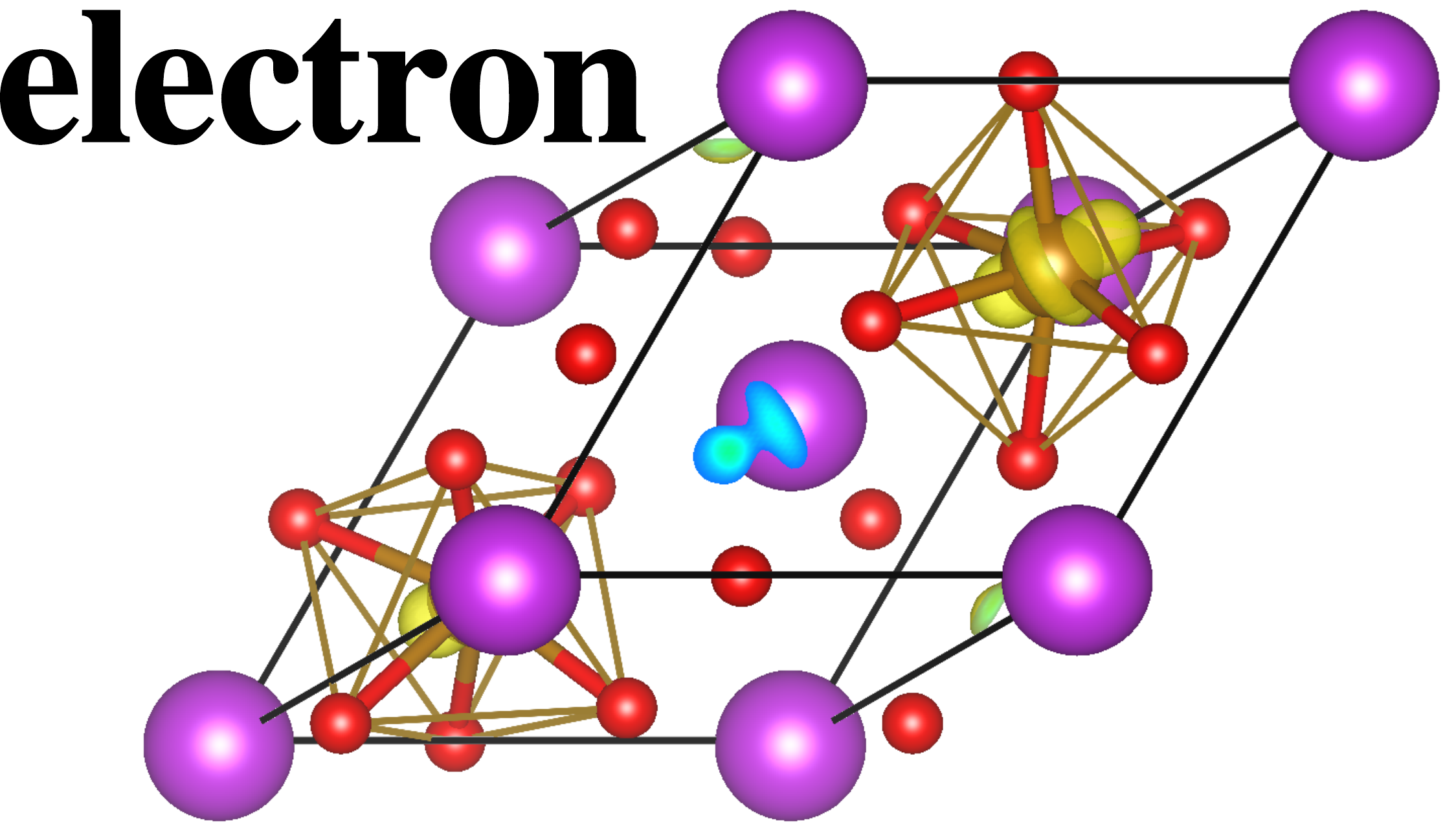}
  \includegraphics[width=0.2\textwidth]{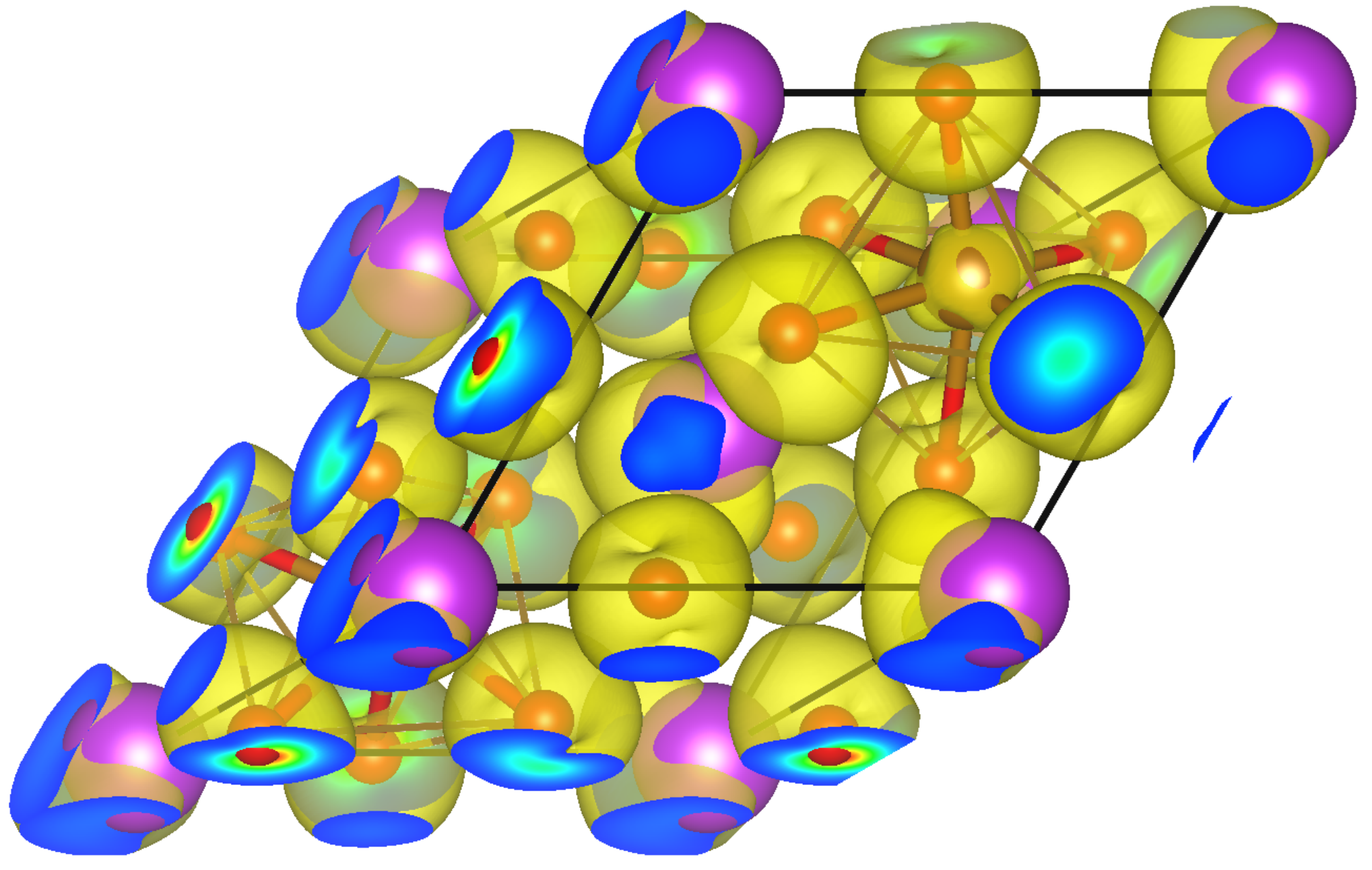}
  \includegraphics[width=0.2\textwidth]{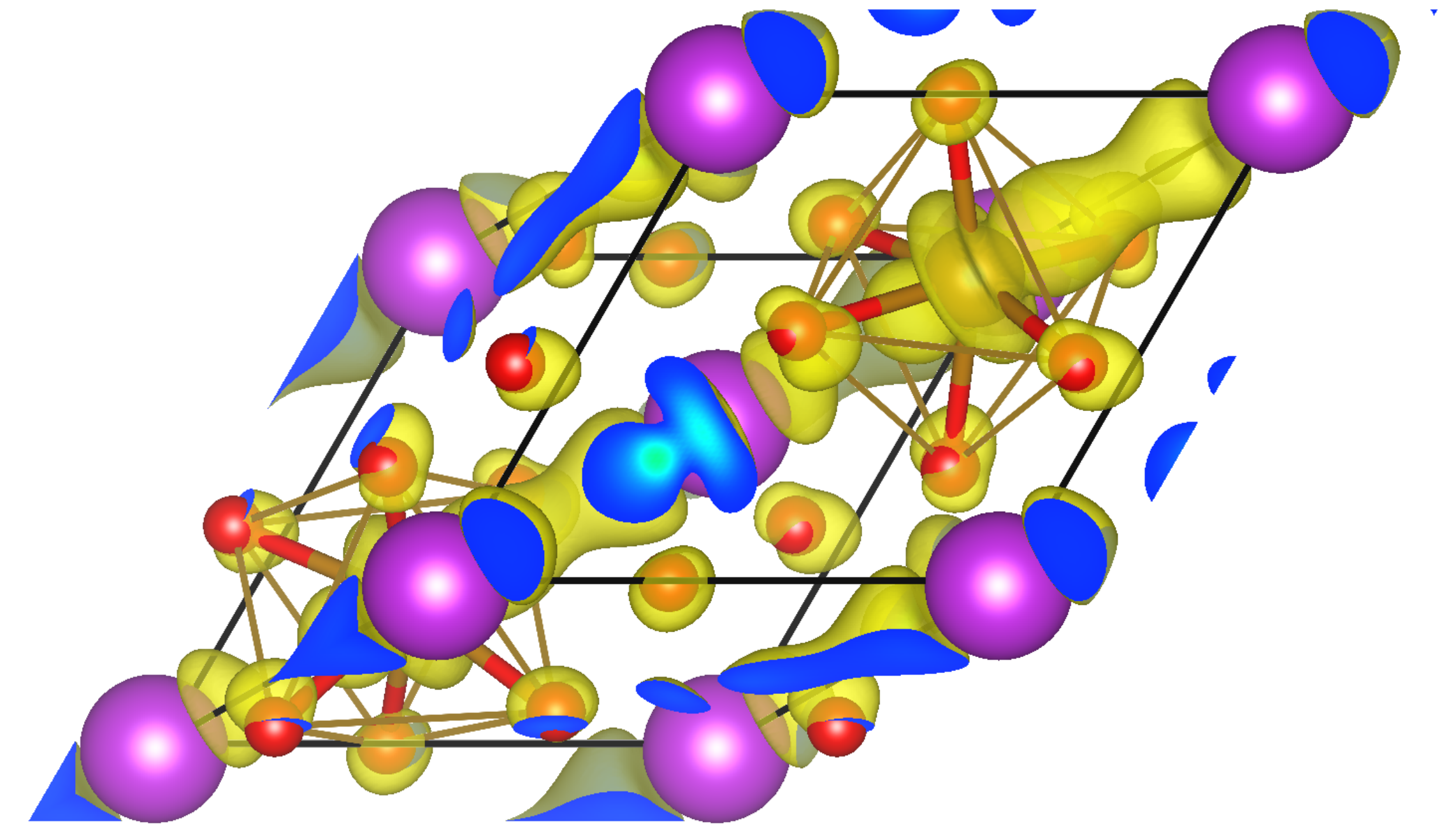}
  \caption{\label{fig:BSE_states}
	Rhombohedral unit cell of BiFeO$_3$ with the isosurfaces (top: 10\% of the maximum; bottom: 3\% and 2\% of the maximum) of the weighted charge densities of the hole (left) and electron (right) states that form the lowest transition in the BSE. 
  }
\end{figure}
The lowest transition from hybridized O-$p$ and Fe-$d$ states to Fe-$d$ states (Fig.~\ref{fig:BSE_states})  should be partially dipole- and spin-forbidden \cite{pisarev:2009:charge}, 
so its oscillator strength would be underestimated in a calculation without spin-orbit coupling, as performed here.
Self-consistency in $GW$, not considered here, might further modify the calculated spectral weight at the onset.

\subsection{Optical spectra of BiFeO$_3$ with defects}
\subsubsection{Absorption}
\begin{figure}
  \includegraphics[width=0.48\textwidth]{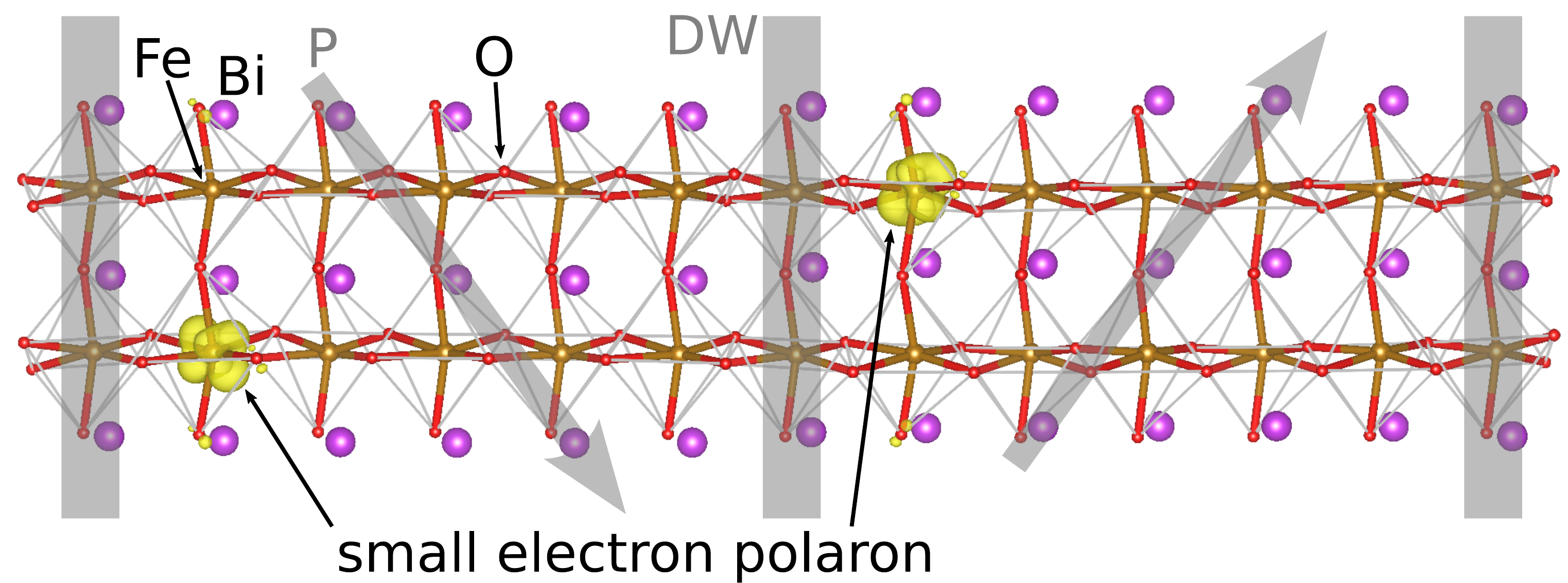}
  \includegraphics[width=0.48\textwidth]{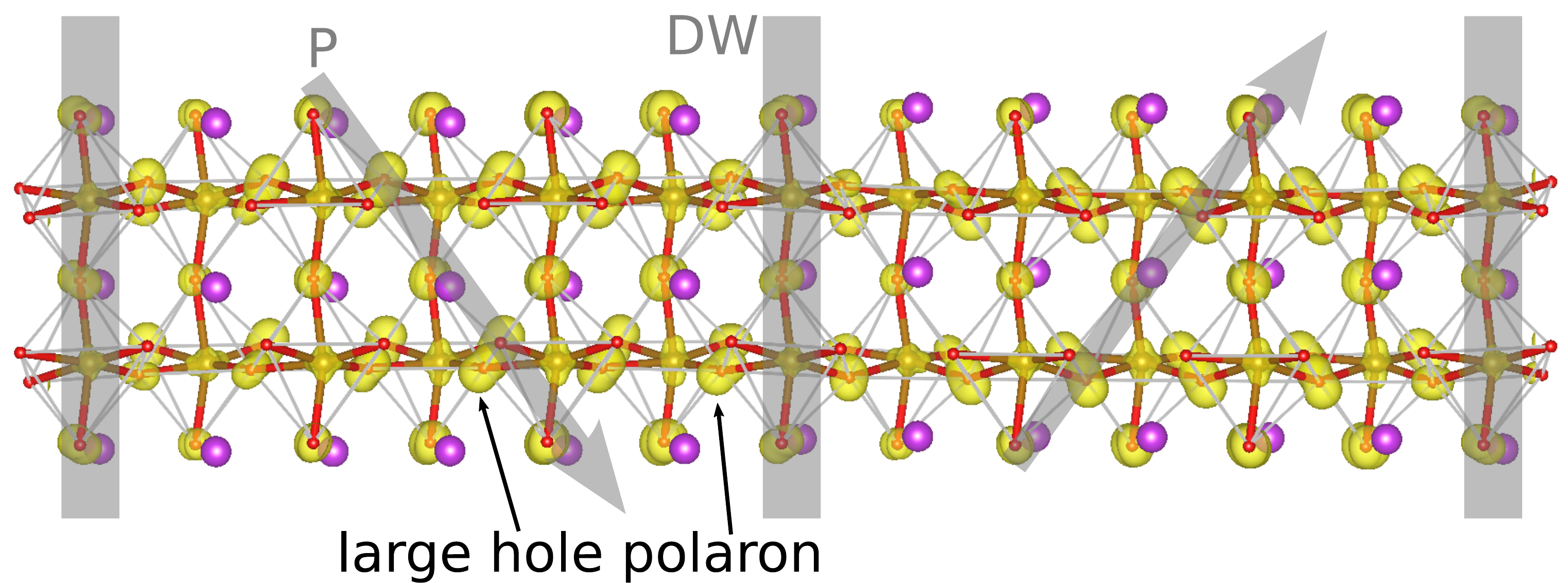}
  \caption{\label{fig:isosurfaces}(Color online) 71\textdegree\ domain wall with small electron polarons (top),
  and large hole polarons (bottom). 
	The yellow objects are charge density isosurfaces of the excess electron (at about 3\% of the maximum density) and hole (at about 14\% of the maximum density), respectively.
  Gray bars and arrows schematically indicate domain walls (DW) and polarization (P) directions, respectively.
  }
\end{figure}

\begin{figure}[htb]
  \includegraphics[width=0.49\textwidth]{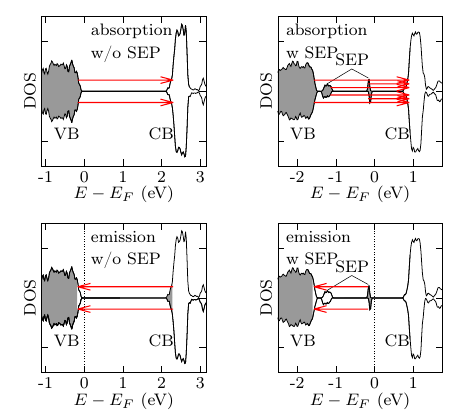}
	\caption{\label{fig:dos_for_eps_and_PL}Transitions involved in absorption and emission (schematic). 
	The transition between higher and lower SEP (small electron polaron) levels is spin-forbidden and not considered here.}
\end{figure}

Optical spectra of BiFeO$_3$ with defects were calculated 
using the independent-particle approximation on the LSDA+$U$ level ($U$=5.3~eV) validated in the previous section.
\paragraph*{Polarons.}
Previous first-principles calculations indicate that excess electrons spontaneously form small polarons \cite{koerbel:2018:electron,radmilovic:2020:combined}, 
whereas holes form large polarons \cite{koerbel:2018:electron,geneste:2019:polarons}.
The large hole polaron occupies a metal-like state at the top of the valence band, 
whereas the small electron polaron occupies a midgap state 
such that the system retains a finite band gap \cite{koerbel:2018:electron,radmilovic:2020:combined}.
Figure~\ref{fig:isosurfaces} shows the atomic configuration of the 71\textdegree\ domain wall 
and isosurfaces of the charge densities of excess electrons and holes.
%
%
Figure~\ref{fig:dos_for_eps_and_PL} schematically shows the transitions involved in absorption and emission for pristine BiFeO$_3$ and BiFeO$_3$ with small electron polarons 
as an example for a defect with a deep level.

\begin{figure}[htb]
  \includegraphics[width=0.49\textwidth]{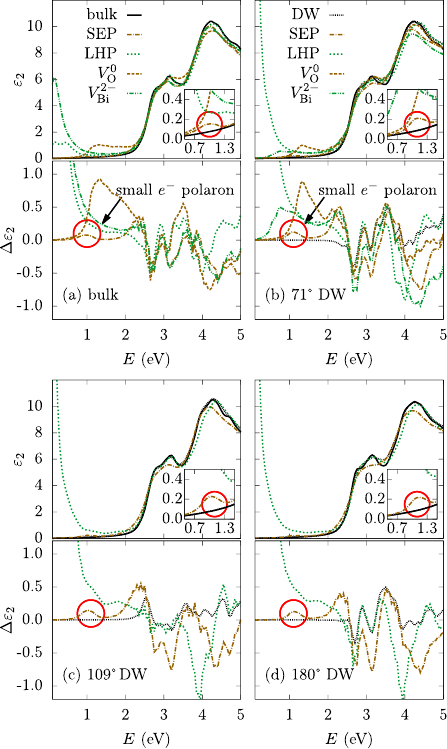}
  \caption{\label{fig:eps_DW} Imaginary part of the microscopically averaged dielectric permittivity, $\varepsilon_2$, 
  of BiFeO$_3$ without (bulk) and with domain walls (DW)
  and with small electron polarons (SEP) and large hole polarons (LHP) at the domain walls.    
  (a) 71\textdegree~wall; (b) 109\textdegree~wall; (c) 180\textdegree~wall.
  The bottom panels show differential spectra with respect to those of the bulk ($\Delta \varepsilon_2=\varepsilon_2-\varepsilon_{2,\mathrm{bulk}}$). 
  Red circles mark peaks from small electron polarons. 
  }
\end{figure}

\begin{figure}[htb]
  \includegraphics[width=0.49\textwidth]{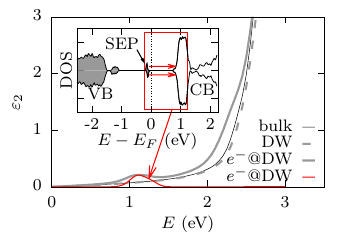}
  \caption{\label{fig:eps2_dos_71DW} (Color online) Imaginary part of the dielectric permittivity $\varepsilon_2$
  for the small electron polaron (SEP) at the 71\textdegree\ wall.
  The inset shows the electronic density of states (DOS) in arbitrary units.
  The red line marks the part of $\varepsilon_2$ that stems only from electronic transitions from the polaron level to the lower conduction bands (red rectangle).
  $E_F$ is the highest occupied level. 
  }
\end{figure}

\begin{figure}
 \includegraphics[width=0.5\textwidth]{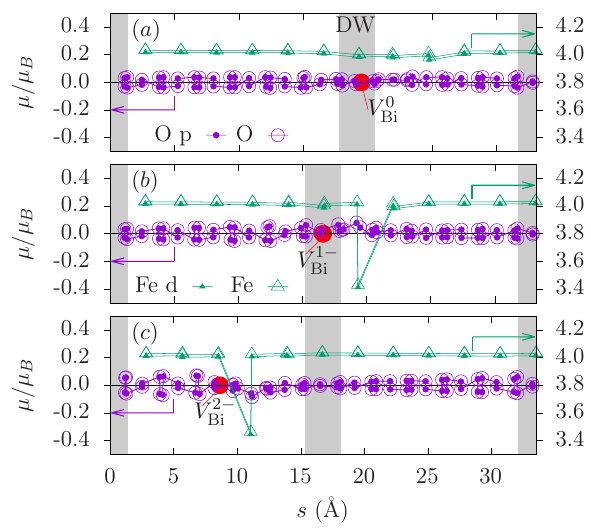}
	\caption{\label{fig:VBi_magmoms}Local magnetic moments of O and Fe in a supercell containing Bi vacancies in different charge states [(a) $V_{\mathrm{Bi}}^{0}$ with three holes, (b) $V_{\mathrm{Bi}}^{-}$ with two holes, (c) $V_{\mathrm{Bi}}^{2-}$ with one hole] near ferroelectric 71\textdegree~domain walls (DWs).}
\end{figure}

Figure~\ref{fig:eps_DW} shows the optical spectra (imaginary part of the high-frequency relative dielectric permittivity $\varepsilon$, $\varepsilon_2$) 
of BiFeO$_3$ without (bulk) and with defects,  
and the difference between them ($\Delta\varepsilon_2$). 
\noindent 
The optical spectrum of the pristine domain wall is nearly identical to that of the bulk.
The large hole polaron leads to large absorption at vanishing photon energy due to its metallic density of states (Drude peak).
In the case of the small electron polaron, 
instead an absorption peak is found at a photon energy of $\approx$1~eV, 
which is well separated in energy from the band edges, so the system remains insulating or semiconducting.
For better visibility of the polaron peaks also the differential spectra 
($\Delta \varepsilon_2=\varepsilon_2-\varepsilon_{2,\mathrm{bulk}}$) are shown in the bottom panels of Fig.~\ref{fig:eps_DW}. 
The small electron polaron peak is composed of electronic transitions between the small electron polaron level and the lowest conduction bands, 
as depicted in Fig.~\ref{fig:eps2_dos_71DW}, which shows $\varepsilon_2$ around the small polaron peak, 
the electronic density of states, and the contribution to $\varepsilon_2$ 
that stems only from electronic transitions 
between the polaron level just below 0 eV and the lowest conduction bands near 1~eV (within the red rectangle).
The electronic density of states of BiFeO$_3$ with small electron polarons is in close agreement with that found in Ref.~[\onlinecite{radmilovic:2020:combined}].
The spectra with oxygen vacancies also contain states within the gap, in agreement with Ref.~[\onlinecite{radmilovic:2020:combined}].
Transitions involving this localized electronic state near an oxygen vacancy appear to have an oscillator strength larger than that of the electron polaron in the bulk.
In the case of bismuth vacancies near domain walls, the calculated local magnetic moments of Fe indicate that holes localize on Fe, forming Fe$^{4+}$, see Fig.~\ref{fig:VBi_magmoms}.
Formation of Fe$^{4+}$ near Bi vacancies at domain walls was also reported in Ref.~[\onlinecite{rojac:2017:domain}] based on spatially resolved electron-energy loss spectroscopy.
\newline
\indent The spectra of the 109\textdegree\ and 180\textdegree\ domain walls are similar 
to that of the 71\textdegree\ wall.
The most pronounced effects of the domain walls are the additional absorption peaks stemming from 
the defects located there.
The oscillator strength of transitions involving defects is already small at the high doping level of the order of 10$^{20}$--10$^{21}$/cm$^3$ 
considered here, and should decrease linearly for lower doping levels.
Therefore, these defect signatures will be difficult to detect in absorption spectroscopy, but should be easier to find using photoluminescence spectroscopy (Kasha's rule). 
\subsubsection{Photoluminescence}
\begin{figure}[htb]
  \includegraphics[width=0.49\textwidth]{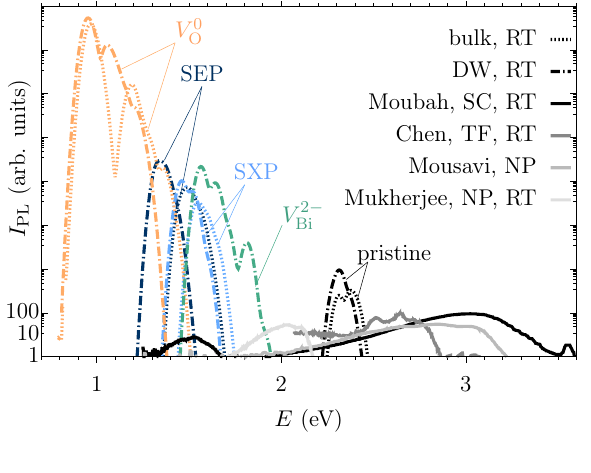}
  \caption{\label{fig:PL_DW} Calculated room-temperature photoluminescence spectra of BiFeO$_3$ without (bulk, dotted lines) 
  and with 71\textdegree~domain walls (DW, dot-dashed lines), without (pristine) and with small electron polarons (SEP, dark blue lines), small exciton polarons (SXP, bright blue lines), oxygen vacancies ($V_{\mathrm{O}}$, orange lines), and Bi vacancies ($V_{\mathrm{Bi}}$, green line).    
	Experimental spectra (solid lines) are from Ref.~[\onlinecite{moubah:2012:photoluminescence}] (the peak at centered at 1.5~eV was ascribed to half-harmonic generation, not to defects), 
	[\onlinecite{chen:2012:optical}], [\onlinecite{mukherjee:2018:visible}], and [\onlinecite{mousavi:2022:synthesis}]. 
	RT: room temperature. SC: Single-domain single crystal. TF: Thin film. NP: nanoparticle.
  }
\end{figure}

Figure~\ref{fig:PL_DW} shows calculated PL spectra of the bulk 
and the 71\textdegree~domain walls without and with defects together with experimental data.
Other than in the case of the absorption spectra, the defect peaks deep inside the gap 
are no longer weak, even though here, other than for the absorption spectra above, 
a low defect concentration of the order of 10$^{11}$/cm$^3$ (excess electrons, oxygen vacancies) and 10$^{15}$/cm$^3$ (Bi vacancies) is assumed. 
The differences between bulk and domain walls and between electron and exciton polarons are subtle.
It appears that the energy levels of the small electron polaron, the small exciton polaron, and the electrons at the oxygen vacancy are slightly lower at the domain wall than in the bulk, 
confirming that the domain walls act as traps for these defects.
Like in the absorption spectra, the oscillator strengths of the transitions of electrons at oxygen vacancies are larger than those for electrons in bulk,
otherwise the properties (energy level and shape of the wave function) of small electron polarons at oxygen vacancies 
\cite{radmilovic:2020:combined}
appear to be quite similar to those of small electron polarons in bulk or at domain walls, indicating that the properties of the small electron polaron 
may be largely independent of the nature of the electron trap.
In the case of the Bi vacancy, it appears that if it is located near a domain wall, it can create a mid-gap level with a separate peak in emission.
The differences in peak position, width, and shape between measured and calculated photoluminescence spectra should arise from neglect of electron-phonon interaction, from the band gap underestimation by LSDA+$U$, and possibly from a non-thermal carrier distribution in experiment. 
Note that the exact shape of the calculated spectra, such as their tail at the onset, is sensitive to the type of broadening that was applied in the calculation (see Fig.~23 in the Supplemental Material \cite{Note1}.) 
We can conclude 
that (a) 
partial thermalization of the excited charge carriers should make it possible to detect the defect levels in photoluminescence, 
and (b) 
electron polarons, electrons at oxygen vacancies, and holes at Bi vacancies located at domain walls may have similar signatures and could be difficult to distinguish.
If both the oxygen vacancy and the isolated electron polaron are present in similar concentrations, the peak of the oxygen vacancy should be dominant.
The predicted photoluminescence spectra could be experimentally verified or falsified by performing photoluminescence spectroscopy of samples that have been prepared to contain a high concentration of domain walls and a specific point defect, 
which could be achieved by means of substrate strain (domain walls) and an oxygen-poor environment (oxygen vacancies) or by ensuring a Bi deficiency during processing (Bi vacancies).

%
%
%
\section{Summary and Conclusions}
\paragraph*{Pristine BiFeO$_3$.}
The optical absorption spectrum of BiFeO$_3$ calculated on the BSE@$G_0W_0$ level, using pseudopotentials and starting from LSDA+$U$, is in close agreement with experiment, 
but different choices of pseudopotentials and different numbers of frequencies in the $GW$ part can change the calculated gap by several tenths of an eV.
A quasiparticle band gap of $\geq$3.1~eV and an optical band gap of $\geq$2.9~eV are obtained (excitonic binding energy $\approx$0.2~eV). 
\newline
\indent Due to partial cancellation of errors, the independent-particle approximation on LSDA+$U$ level has a similar accuracy.
The agreement with experiment is best for $U\approx6$~eV.
Assuming that the methodology adopted here is valid, the shoulder at
the absorption onset cannot be explained by defect states, such as
those at oxygen vacancies, because the defect states lie too deep in
the gap so they form separate peaks rather than a shoulder.
Moreover, the defect concentration would need to be quite large (of the order of $10^{20}$/cm$^3$) 
 to give rise to a sizable signal in the absorption spectrum.
The shoulder is already present in calculated absorption spectra of defect-free BiFeO$_3$.
It seems, therefore, more likely that the shoulder is composed of transitions between hybridized O-$p$/Fe-$d$ valence states and Fe-$d$ conduction states 
of pristine BiFeO$_3$. Such transitions that are partially dipole- and spin-forbidden are likely underestimated by the methodology adopted here.
To validate the adopted methodology and to clarify the origin of the shoulder unambigously,
self-consistency in $GW$ and a dense $k$-point mesh using e.g. interpolation should be employed, 
taking into account spin-orbit coupling, ideally also avoiding the use of pseudopotentials, a formidable task.

\paragraph*{BiFeO$_3$ with defects.}
Different point defects, namely, electron polarons, electrons at oxygen vacancies, and possibly holes near bismuth vacancies apparently introduce defect states deep within the gap. 
These defect states seem to be at similar energies and may be difficult to distinguish.
In the case of hole doping ($p$-type conditions), a Drude peak should appear due to the metal-like density of states of the large hole polaron.
The spectroscopic signatures of the point defects should only be visible in absorption spectra at high defect concentrations of the order of 10$^{20}$/cm$^3$,
whereas they should be visible in photoluminescence at lower concentrations, such as 10$^{11}$/cm$^3$, provided the excited carriers are near thermal equilibrium and their concentration is of the order of the defect concentration or lower, otherwise bulk transitions should dominate.
The ferroelectric domain walls seem to accumulate these point defects rather than to affect the spectra directly.
\section*{Acknowledgement}
S.~K. thanks F. Bechstedt, S. Botti, J. Hlinka, and S. Sanvito for helpful discussion, 
P. Ghosez for pointing out literature, and S. Choi for sharing ellipsometry data.
This project  has  received  funding  from  the  
European Union's Horizon 2020 research and innovation programme under the Marie Sk\l{}odowska-Curie 
Grant Agreement No. 746964.
Computational resources and support were supplied by the Trinity Centre for High Performance Computing funded by Science Foundation Ireland,
by the Irish Centre for High-End Computing, and 
and by the e-INFRA CZ project (ID:90254), supported by the Ministry of Education, Youth and Sports of the Czech Republic,
and the HPC cluster ARA of the University of Jena, Germany.
Figures were made using {\sc{Vesta}} \cite{momma:2011:vesta} and gnuplot.


\bibliography{jabbr,all}
\begin{figure*}[h]
	\Huge{Supplemental Information\\
	Optical signatures of defects in BiFeO$_3$}
\end{figure*}
\clearpage
\setcounter{section}{0}
%
%
%
%
%
%




\section{Convergence tests for bulk BiFeO$_3$}
This section contains convergence tests for the 10-atom unit cell of perfect rhombohedral BiFeO$_3$. 
Details of the pseudopotentials (PSP) used and calculation results are given in Table~\ref{tab:psp_results}.
Where not stated otherwise, test calculations were carried out with a plane-wave cutoff energy $E_{\mathrm{cut}}=520~$eV,
a cutoff energy for the dielectric permittivity $E_{\mathrm{cut}}(\varepsilon)$=300~eV, and a $2\times 2\times 2$ $k$-point grid,
$\geq$1560 bands in total, of which $\geq 80$ bands were corrected with $G_0W_0$,
neglecting spin-orbit coupling.
%
\subsection{\label{sec:kpoints}$k$-points}

\begin{figure}[htb]
  \includegraphics[width=0.45\textwidth]{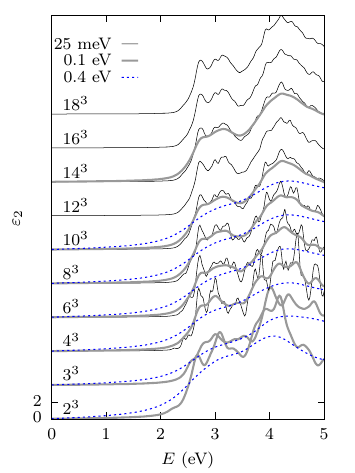}
	\caption{\label{fig:eps_vs_kpts_IP} Imaginary part of the dielectric permittivity 
	$\varepsilon_2$ calculated on DFT-IPA level as a function of $k$-point density ($2\times 2\times 2$, $2^3$ to $18\times 18\times 18$, $18^3$ $k$-points) and spectral broadening (25~meV to 0.4~eV).
  }
\end{figure}

Figure~\ref{fig:eps_vs_kpts_IP} shows $\varepsilon_2$ obtained in the independent particle approximation (IPA) on LSDA+$U$ level ($U$=5.3~eV)
using PSP0
for different $k$-point meshes ($2\times 2\times2$ to $18\times18\times 18$) and different spectral broadenings (25~meV to 0.4~eV). 
The $k$-point convergence depends on the broadening: For 25~meV a dense mesh of $\approx$ $12\times 12\times 12$ $k$-points is needed,
for 0.4~eV a mesh of $\approx3\times 3\times 3$ $k$-points already yields nearly converged spectra.

The $k$-point convergence of the BSE spectra is shown in section~\ref{sec:model_BSE}. 
\subsection{\label{sec:nomega}Number of frequencies in $GW$}
\begin{figure*}[htb]
  \includegraphics[width=0.49\textwidth]{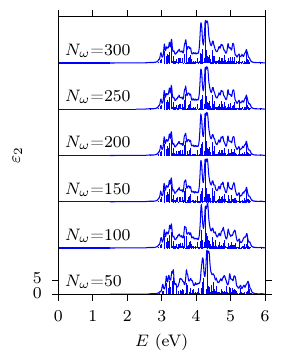}
  \includegraphics[width=0.49\textwidth]{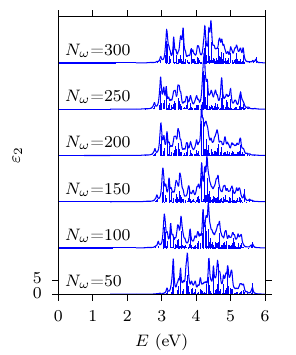}
	\caption{\label{fig:eps_nomega_GW_psp} Imaginary part of the dielectric permittivity 
    $\varepsilon_2$ for light polarization along the ordinary and extraordinary axis (0.2~eV broadening) and oscillator strength calculated on $G_0W_0$+BSE level as a function of the number of frequencies $N_{\omega}$ included in the $GW$ step using PSP1 (left) and PSP2 (right).
  }
\end{figure*}

\begin{figure*}[htb]
  \includegraphics[width=0.49\textwidth]{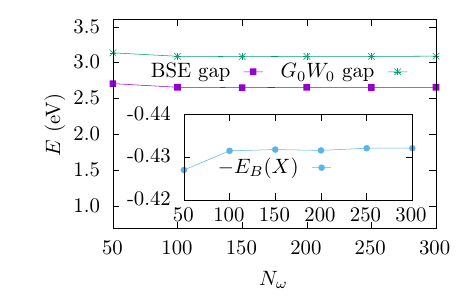}
  \includegraphics[width=0.49\textwidth]{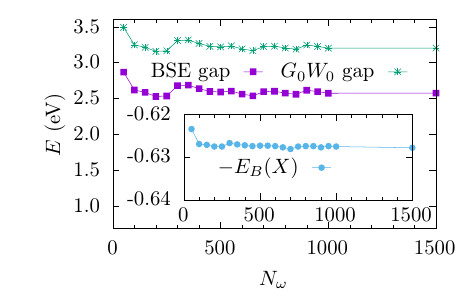}
	\caption{\label{fig:E_B_X_nomega} $G_0W_0$ \& BSE@$G_0W_0$ gaps, and the exciton binding energy $E_B(X)$ as a function of $N_{\omega}$ using PSP1 (left) and PSP2 (right).
  }
\end{figure*}

Figure~\ref{fig:eps_nomega_GW_psp} shows $\varepsilon_2$ (broadened by 0.1~eV) and the oscillator strengths from the BSE@$G_0W_0$ as a function of the number of frequencies $N_{\omega}$ in $GW$,
using PSP1 and PSP2, $U$=6~eV, and 16 valence and 16 conduction bands in the BSE.
Figure~\ref{fig:E_B_X_nomega} shows the gaps from $GW$ and the BSE as well as the exciton  binding energy.
The exciton binding energy and the spectral shape are approximately converged at $N_{\omega}\approx$100.
The gaps themselves converge more slowly and still vary by $\approx$0.1~eV above $N_{\omega}=$100.

\subsection{\label{sec:nbands}Total number of bands}
\begin{figure}[htb]
  \includegraphics[width=0.49\textwidth]{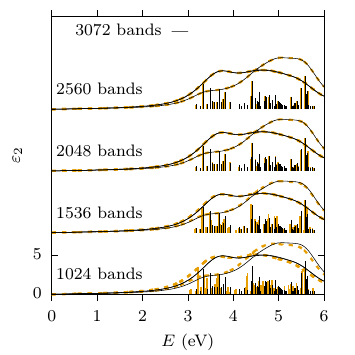}
	\caption{\label{fig:eps_nbands_GW} Imaginary part of the dielectric permittivity 
	$\varepsilon_2$ and oscillator strength calculated on BSE@$G_0W_0$ level as a function of the total number of bands (39 bands are occupied; 80 bands are corrected with $G_0W_0$). 
	Shown are both curves for light polarization along the ordinary and extraordinary axis (dashed yellow curves) in comparison with the spectrum for 3072 $GW$ bands (solid black line).
  }
\end{figure}

Figure~\ref{fig:eps_nbands_GW} shows $\varepsilon_2$ (broadened by 0.4~eV) and the oscillator strengths from the BSE@$G_0W_0$ as a function of the total number of bands included 
in the $GW$ part using $E_{\mathrm{cut}}(\varepsilon)=266.7$~eV, PSP0, $U=5.3$~eV, $N_{\omega}=75$, and 8 valence and 8 conduction bands in the BSE.
Using 1560 bands all transition energies up to 5.7~eV are converged within 40~meV compared to the calculation with 3072 bands.  
For 2048 bands all transition energies up to 6~eV are converged within 3~meV.

\subsection{\label{sec:nbands_bse}Number of bands in the BSE}
\begin{figure}[htb]
  \includegraphics[width=0.49\textwidth]{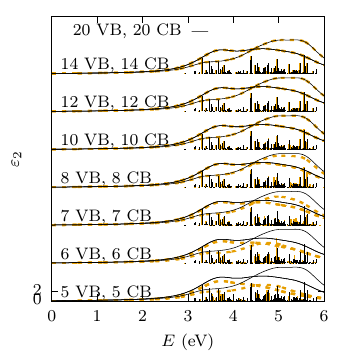}
	\caption{\label{fig:eps_nbands_BSE} Imaginary part of the dielectric permittivity 
	$\varepsilon_2$ and oscillator strength calculated on BSE@$G_0W_0$ level as a function of the number of valence (``VB'') and conduction (``CB'') bands included in the BSE part. 
	Shown are both curves for light polarization along the ordinary and extraordinary axis in comparison with a converged spectrum (20 VB, 20 CB).
  }
\end{figure}

Figure~\ref{fig:eps_nbands_BSE} shows $\varepsilon_2$ (broadened by 0.4~eV) and the oscillator strengths from the BSE@$G_0W_0$ as a function of the number of valence and conduction bands (VB and CB) included 
in the BSE part, using $U$=5.3~eV, PSP0, $N_{\omega}=75$, and $E_{\mathrm{cut}}(\varepsilon)=266.7$~eV. The spectra obtained with 12 VB and 12 CB are already identical to those obtained with 20 VB and 20 CB. 

\subsection{\label{sec:hubbard_u}Hubbard $U$}
\begin{figure}
  \includegraphics[width=0.49\textwidth]{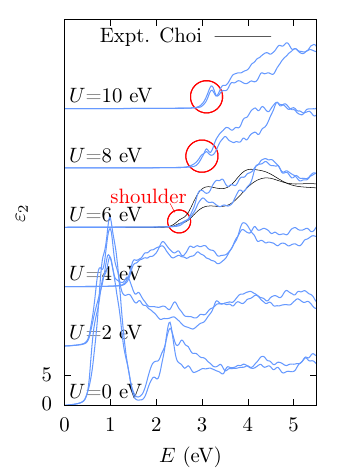}
	\caption{\label{fig:eps_vs_U}$\varepsilon_2$ as a function of the Hubbard $U$. For $U$=6~eV the spectrum is close to the experimental one of Ref.~\cite{choi:2011:optical}, including the shoulder. The broadening is 25~meV.}
\end{figure}

\begin{figure}[htb]
  \includegraphics[width=0.49\textwidth]{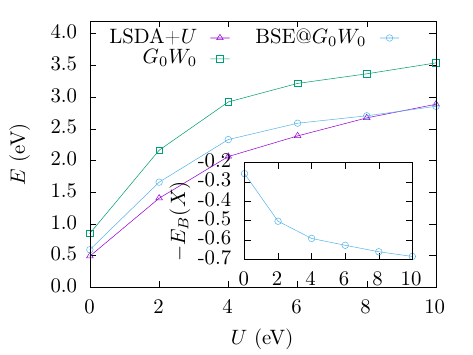}
	\caption{\label{fig:gaps_vs_U}  
	Kohn-Sham (KS), $G_0W_0$, and BSE@$G_0W_0$ gaps as a function of the Hubbard parameter $U$ for PSP2. The inset shows the exciton binding energy $E_B(X)$.
  }
\end{figure}

Figure~\ref{fig:eps_vs_U} shows $\varepsilon_2$ (broadened by 25~meV) on LSDA+$U$ level as a function of the Hubbard $U$, using PSP2, 120 bands, and a $10\times 10\times 10$ $k$-point grid. 
The best agreement with experiment is obtained for $U\approx$6~eV.
Figure~\ref{fig:gaps_vs_U} shows the Kohn-Sham (LSDA+$U$), $G_0W_0$, and BSE@$G_0W_0$ gaps as a function of the Hubbard $U$, using PSP2, a $2\times 2\times 2$ $k$-point grid, 
$N_{\omega}=150$, and 16 valence and 16 conduction bands in the BSE.
Both the gaps and the exciton binding energy depend considerably on the starting point, {\textit{i. e.}} the $U$ parameter.

\subsection{\label{sec:ecut}Plane wave energy cutoff}
\begin{figure*}[htb]
  \includegraphics[width=0.49\textwidth]{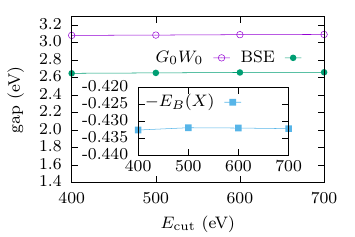}
  \includegraphics[width=0.49\textwidth]{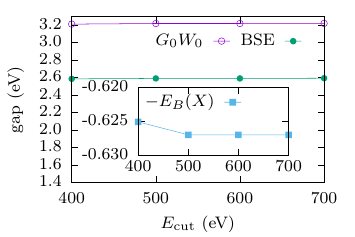}
    \caption{\label{fig:gaps_different_ENCUT} $G_0W_0$ and BSE@$G_0W_0$ gaps and exciton binding energies $E_B(X)$ 
    vs. plane wave cutoff energy $E_{\mathrm{cut}}$ using PSP1 (left) and PSP2 (right).
  }
\end{figure*}

Fig.~\ref{fig:gaps_different_ENCUT} shows the $G_0W_0$ and BSE@$G_0W_0$ gaps and the exciton binding energy $E_B(X)$ as a function of the cutoff energy of the basis functions $E_{\mathrm{cut}}$, 
using PSP1 and PSP2, $U$=6~eV, $N_{\omega}=150$, $E_{\mathrm{cut}}(\varepsilon)=300$~eV, and 16 valence and 16 conduction bands in the BSE.
The gaps and exciton binding energies are converged at $E_{\mathrm{cut}}\approx 500$~eV.

\subsection{\label{sec:ecut_eps}Energy cutoff for $\varepsilon$}
\begin{figure*}[htb]
  \includegraphics[width=0.49\textwidth]{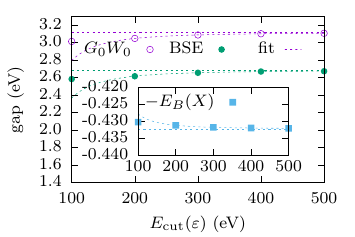}
  \includegraphics[width=0.49\textwidth]{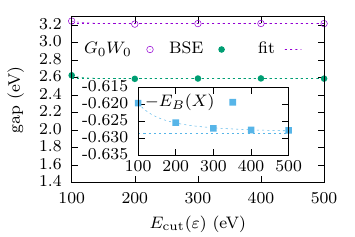}
	\caption{\label{fig:gaps_different_ENCUTGW} $G_0W_0$ and BSE@$G_0W_0$ gaps and exciton binding energies $E_B(X)$ vs. cutoff energy of the dielectric matrix $E_{\mathrm{cut}}(\varepsilon)$
     using PSP1 (left) and PSP2 (right).
	The horizontal dashed lines mark the extrapolated gaps from a fitted power law.
  }
\end{figure*}

Fig.~\ref{fig:gaps_different_ENCUTGW} shows the $G_0W_0$ and BSE@$G_0W_0$ gaps and the exciton binding energy $E_B(X)$ as a function of the cutoff energy of the dielectric matrix $E_{\mathrm{cut}}(\varepsilon)$,
using PSP1 and PSP2, $U$=6~eV, $N_{\omega}=150$, and 16 valence and 16 conduction bands in the BSE.
The gaps are converged at  $E_{\mathrm{cut}}(\varepsilon)\approx 500$~eV (PSP1) and $\approx 200$~eV (PSP2), respectively.

\begin{figure*}[htb]
  \includegraphics[width=0.49\textwidth]{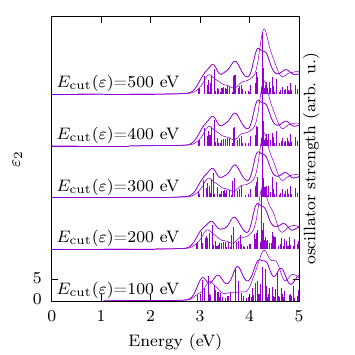}
  \includegraphics[width=0.49\textwidth]{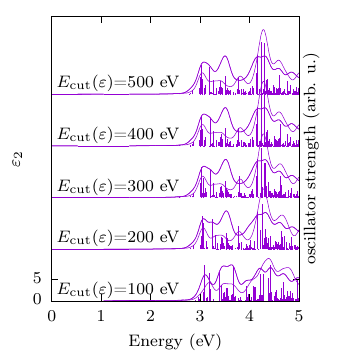}
    \caption{\label{fig:BSE_EPS2_and_EV_different_ENCUTGW} $\varepsilon_2$ and oscillator strengths from BSE@$G_0W_0$ as a function of the cutoff energy of the dielectric matrix $E_{\mathrm{cut}}(\varepsilon)$, using PSP1 (left) and PSP2 (right).
  }
\end{figure*}

Fig.~\ref{fig:BSE_EPS2_and_EV_different_ENCUTGW} shows $\varepsilon_2$ (broadened by 0.1~eV) and the oscillator strengths for different $E_{\mathrm{cut}}(\varepsilon)$ 
using PSP1 and PSP2.
The shape of the spectra and the oscillator strengths of the transitions near the onset are converged at $E_{\mathrm{cut}}(\varepsilon))\approx$200~eV.

\subsection{\label{sec:model_BSE}model BSE}
\begin{figure*}
 \includegraphics[width=0.4\textwidth]{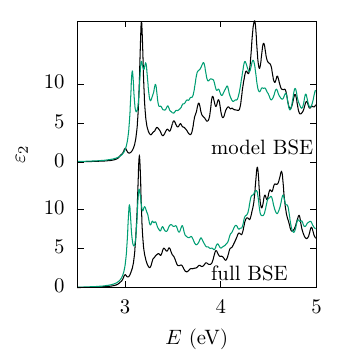}
 \includegraphics[width=0.4\textwidth]{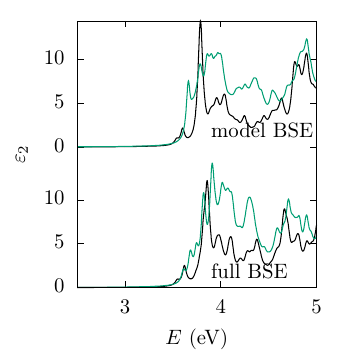}
	\caption{\label{fig:eps2_GW_BSE_model_BSE} 
	$\varepsilon_2$ 
	from the BSE@$G_0W_0$ and the model BSE for $6\times 6\times 6$ $k$-points (top, broadened by 25~meV)
    and BSE spectra and oscillator strengths of the lowest 150 transitions as a function of $k$-points (bottom), using PSP1 (left) and PSP2 (right). SOC was neglected.
	}
\end{figure*}

\begin{figure*}
 \includegraphics[width=0.4\textwidth]{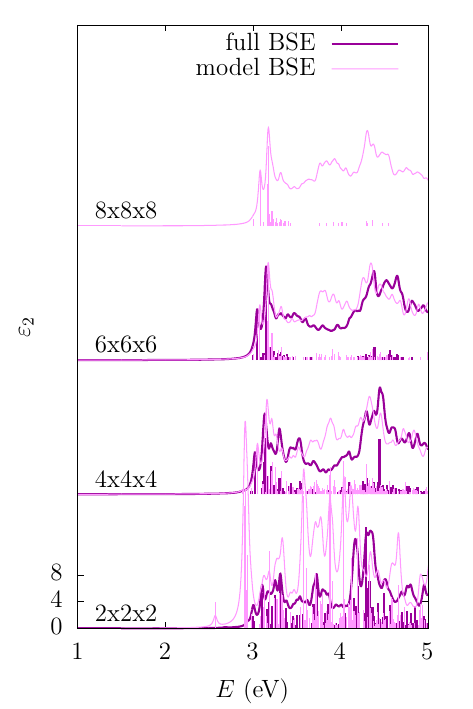}
 \includegraphics[width=0.4\textwidth]{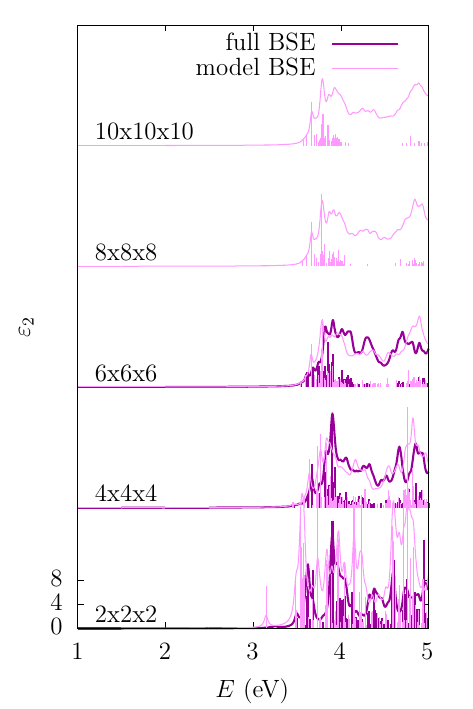}
	\caption{\label{fig:eps2_GW_BSE_model_BSE_kpoints} 
	$\varepsilon_2$ 
	from the BSE@$G_0W_0$ and the model BSE for $6\times 6\times 6$ $k$-points (top, broadened by 25~meV)
    and BSE spectra and oscillator strengths of the lowest 150 transitions as a function of $k$-points (bottom), using PSP1 (left) and PSP2 (right). SOC was neglected.
	}
\end{figure*}

\begin{figure*}
 \includegraphics[width=0.7\textwidth]{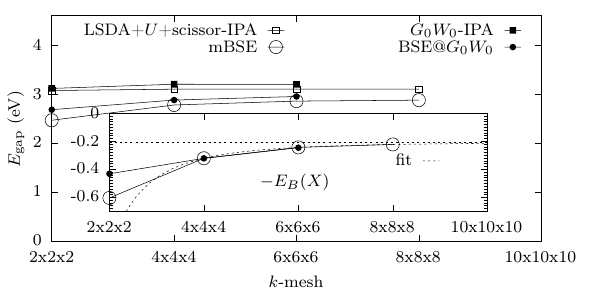}
 \includegraphics[width=0.7\textwidth]{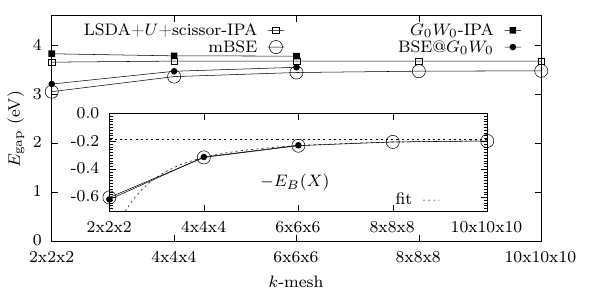}
    \caption{\label{fig:gaps_model_BSE_potcars_Bi_d_Fe_O} Gaps from the model BSE and the full BSE as a function of $k$-points, using PSP1 (left) and PSP2 (right). 
	The inset shows the exciton binding energy.
	The horizontal line marks the binding energy extrapolated to an infinitely dense $k$-point mesh by a fitted power law.}
\end{figure*}

Figure~\ref{fig:eps2_GW_BSE_model_BSE} shows $\varepsilon_2$ from the model BSE (mBSE@LSDA+$U$+scissor) as a function of the $k$-point mesh in comparison with the full BSE@$G_0W_0$ obtained with PSP1 and PSP2, using $U$=6~eV and a $6\times 6\times 6$ $k$-point grid. 
For the full BSE $N_{\omega}=150$ and $E_{\mathrm{cut}}(\varepsilon)=300$~eV were used. 
Both the model and the full BSE were solved for 16 valence and 16 conduction bands.
Figure~\ref{fig:eps2_GW_BSE_model_BSE_kpoints} shows the convergence of full and model BSE with $k$-points. 
Figure \ref{fig:gaps_model_BSE_potcars_Bi_d_Fe_O} shows the gaps obtained from the model BSE in comparison with BSE@$G_0W_0$.
%
%

\subsection{\label{sec:SOC}Spin-orbit coupling}

\begin{figure}[htb]
  \includegraphics[width=0.49\textwidth]{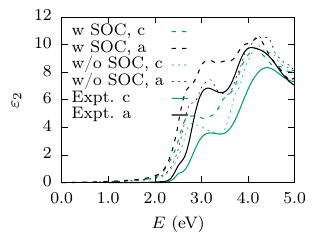}
	\caption{\label{fig:eps_w_and_wo_SOC} Imaginary part of the dielectric permittivity 
	$\varepsilon_2$ on LSDA+$U$ level with and without considering spin-orbit coupling (SOC). Experimental data are from Ref.~\cite{choi:2011:optical}.
  }
\end{figure}

Figure~\ref{fig:eps_w_and_wo_SOC} shows $\varepsilon_2$ (broadened by 0.1~eV) on LSDA+$U$ level obtained with and without including spin-orbit coupling (SOC),
 using PSP0, $U$=5.3~eV, 100 bands, and an $8\times 8\times 8$ $k$-point grid.
SOC has a considerable effect on the optical absorption spectra and should therefore be included in the calculation. 
However, since inclusion of SOC doubles the computation time, reduces the band gap only by $\approx$0.1~eV, 
and does not clearly improve the agreement with experimental spectra, it is neglected in the supercell calculations with defects. 
Neglecting spin-orbit coupling partially cancels the band gap underestimation of LSDA+$U$ with $U$=5.3~eV.


\subsection{\label{sec:LFE}Local field effects}

\begin{figure}[htb]
  \includegraphics[width=0.49\textwidth]{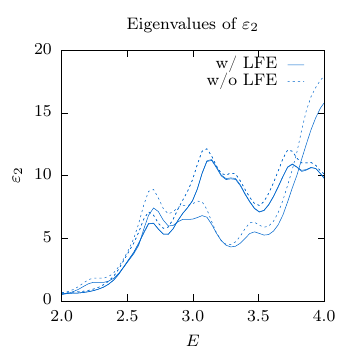}
	\caption{\label{fig:eps_w_and_wo_LFE} Imaginary part of the dielectric permittivity 
	$\varepsilon_2$ on LSDA+$U$ level  with and without considering local field effects (LFE).
  }
\end{figure}

Figure~\ref{fig:eps_w_and_wo_LFE} shows $\varepsilon_2$ (broadened by 0.1~eV) on LSDA+$U$ level ($U$=5.3~eV) obtained with and without local field effects (LFE),
 using PSP0 and 50 bands. 
The local field effects reduce the spectral weight for energies up to 5~eV by about 10\%. 

\subsection{\label{sec:Bi_semicore}Bi semicore electrons}

\begin{figure}[htb]
  \includegraphics[width=0.49\textwidth]{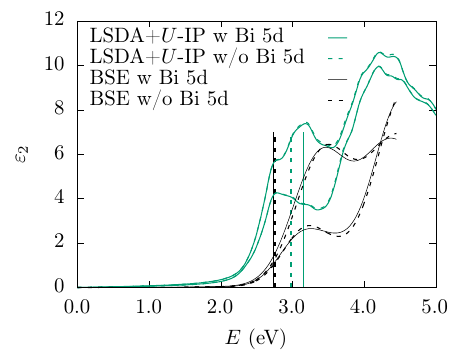}
	\caption{\label{fig:eps_w_and_wo_Bi_semicore} Imaginary part of the dielectric permittivity 
	$\varepsilon_2$ with and without considering Bi semicore (5d) electrons as valence electrons, for light polarizations along the extraordinary (``c'') and ordinary (``a'') axis of rhombohedral BiFeO$_3$.
	Vertical lines mark $G_0W_0$ (green) and BSE gaps (black).
  }
\end{figure}

Figure~\ref{fig:eps_w_and_wo_Bi_semicore} shows $\varepsilon_2$ obtained with and without including Bi semicore electrons in the valence,
using PSP0 (with ``Bi-d'' instead of ``Bi'' in the case of semicore included), $U$=5.3~eV, and 100 bands.
The spectra for light polarization along the principle axes were calculated with LSDA$+U$ on the independent-particle level, neglecting local-field effects, 
for an $8\times 8\times 8$  $k$-point mesh and a broadening of 0.1~eV, and with the BSE@$G_0W_0$ on a $2\times 2\times 2$ $k$-point mesh, broadened by 0.4~eV.
The contribution of the $5d$ electrons of Bi to the optical absorption spectra on LSDA+$U$ level near the absorption onset is negligible, 
but the $G_0W_0$ gap is affected by the Bi semicore electrons.

\subsection{pseudopotentials}
\begin{figure}[htbp]
  \includegraphics[width=0.49\textwidth]{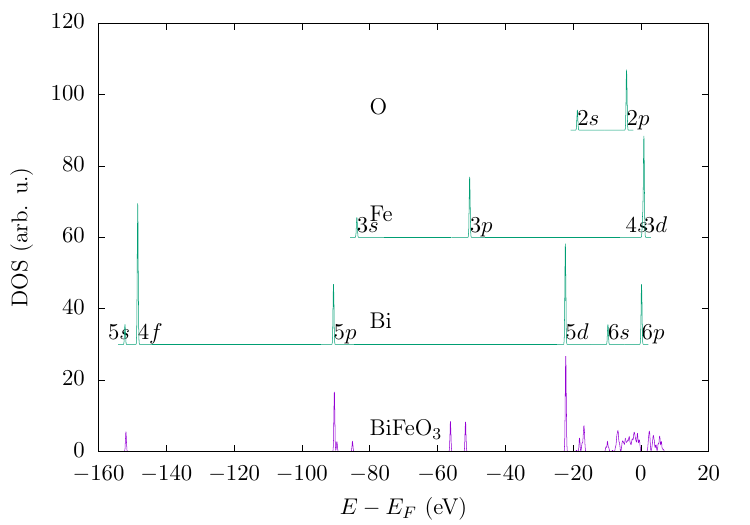}
	\caption{\label{fig:atomic_states} Atomic states involved in the electronic structure of BiFeO$_3$. 
  }
\end{figure}

Figure~\ref{fig:atomic_states} shows the atomic states involved in the electronic structure of BiFeO$_3$. 
The $4f$ states of Bi are missing in the depicted density of states of BiFeO$_3$ because in the pseudopotential they are part of the core electrons. 
Two pseudopotentials with energetically well-separated core and valence states were finally chosen for the $G_0W_0$ calculations, 
namely the ``Bi-d Fe O'' (15, 8, and 6 valence electrons) and the ``Bi-sv-GW, Fe-sv-GW, O-GW'' (23, 16, and 6 valence electrons) sets.

\begin{figure}[htbp]
  \includegraphics[width=0.49\textwidth]{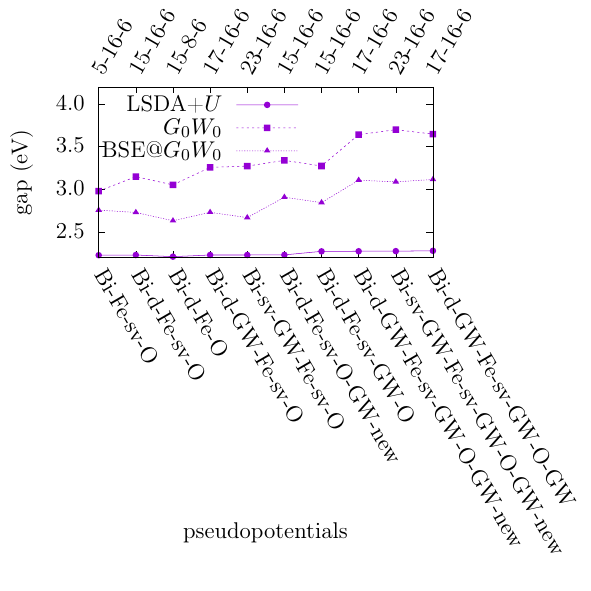}
	\caption{\label{fig:gaps_vs_pspot} Band gap from LSDA+$U$ ($U$=5.3~eV), $G_0W_0$, and the BSE@$G_0W_0$ for different pseudopotentials. 
	The number of included valence electrons for each species (example: 5-16-6 for 5 Bi electrons, 15 Fe electrons, and 6 O electrons) is indicated on the upper abscissa.
	Only 15-16-6 is a configuration with closed valence shells.
  }
\end{figure}

\begin{figure}[htbp]
  \includegraphics[width=0.49\textwidth]{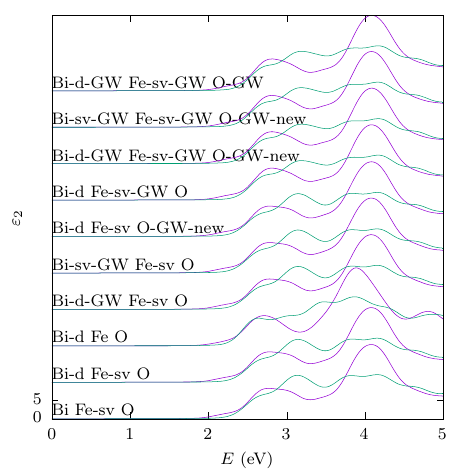}
	\caption{\label{fig:eps_GS2_vs_pspot} The two different eigenvalues of $\varepsilon_2$ on LSDA+$U$ level  ($U$=5.3~eV) for different pseudopotentials, broadened by 0.2~eV. 
  }
\end{figure}

Different pseudopotentials yield different gaps, especially on the $G_0W_0$ level. The BSE gaps are less sensitive to the pseudopotential. 
Figure~\ref{fig:gaps_vs_pspot} shows the gaps on LSDA+$U$ ($U$=5.3~eV), $G_0W_0$, and BSE@$G_0W_0$ level for different pseudopotentials,
using $N_{\omega}$=150.
The $G_0W_0$ gaps vary between 3.0 and 3.6~eV, the BSE gaps between 2.6~and 3.1~eV, the exciton binding energy between 0.2~and~0.6~eV.
Within the pseudopotential approach, as adopted here, it appears difficult to obtain more than a lower limit for the gap.  
Figure~\ref{fig:eps_GS2_vs_pspot} shows $\varepsilon_2$ on LSDA+$U$ level ($U$=5.3~eV, 0.2~eV broadening) obtained with different pseudopotentials. 
Except for the pseudopotential without Fe semicore (``Bi-d-Fe-O'') all spectra are nearly identical on LSDA+$U$ level. 

\begin{figure}[htbp]
  \includegraphics[width=0.49\textwidth]{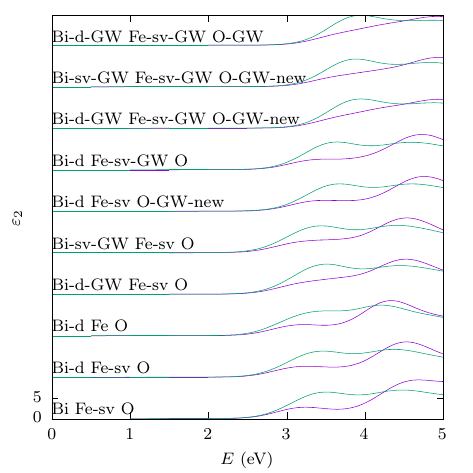}
	\caption{\label{fig:eps_BSE_vs_pspot} The two different eigenvalues of $\varepsilon_2$ from the BSE@$G_0W_0$ for different pseudopotentials, broadened by 0.4~eV. 
  }
\end{figure}

Figure~\ref{fig:eps_BSE_vs_pspot} shows $\varepsilon_2$ (broadened by 0.4~eV) from the BSE@$G_0W_0$ obtained with different pseudopotentials. 
The spectrum onset and the shape of the extraordinary component vary with the pseudopotential.
\begin{table*}
	\begin{tabular}{l|l|l|l}
		                                        & ``Bi Fe-sv O''  (PSP0)                 & ``Bi-d Fe O'' (PSP1)                  & ``Bi-sv-GW Fe-sv-GW O-GW''   (PSP2)               \\\hline
		valence conf.                           &  Bi $5f^0$ $6s^{2}$ $6p^3$ $6d^0$      & Bi $5d^{10}$ $5f^0$  $6s^2$ $6p^3$    & Bi $5s^2$ $5p^6$ $5d^{10}$ $5f^0$ $6s^2$ $6p^3$   \\
				                                &  Fe $3s^2$ $3p^6$ $3d^7$ $4s^1$ $4f^0$ & Fe $3d^{7}$ $4s^1$ $4p^0$ $4f^0$      & Fe $3s^2$ $3p^6$ $3d^8$ $4f^0$                    \\
				                                &  O $2s^2$ $2p^4$ $3d^0$                & O $2s^2$ $2p^4$ $3d^{0}$              & O $2s^2$ $2p^4$ $3d^0$                            \\        
    LSDA+$U$ gap w/o SOC                        &  dir. 2.35, fund. 2.30                 &	dir. 2.36 , fund. 2.36               & dir. 2.41, fund. 2.36                             \\
    LSDA+$U$ gap w SOC                          &  dir. 2.25, fund. 2.19                 & dir. 2.25, fund. 2.25                 &  dir. 2.31, fund. 2.26                            \\ 
		corr. (SOC)                             &  dir. -0.11, fund. -0.11               &         dir. -0.10, fund.  -0.11      &     dir. -0.10, fund.  -0.11                      \\
		corr ($E_{\mathrm{cut}}$($\varepsilon$))&                                        &  +0.03                                &  0.00                                             \\ 
		corr. (BSE $k$-points)                  &                                        &  0.04                                 &     0.05                                          \\ 
		dir. $GW$ gap (uncorr.)                 &                                        &  3.13 (3.20)                          &   3.67 (3.77)                                     \\ 
		fund. $GW$ gap (uncorr.)                &                                        &   3.08 (3.16)                         &  3.57 (3.68)                                      \\ 
		BSE gap (uncorr.)                       &                                        &  2.92  (2.95)                         & 3.5  (3.54)                                       \\ 
		$E_B(X)$                                &                                        &  0.21                               &    0.21                                           \\ 
		scissor                                 &                                        &  0.84 eV                              &      1.36 eV                                      \\ 
		$\varepsilon_{\infty}$	                &                                        &     7.26                              & 7.25                                              \\ 
		$\lambda$                               &                                        &       1.417                           & 1.470                                             \\ 
	\end{tabular}
    \caption{\label{tab:psp_results}Parameters of and properties obtained with different pseudopotentials}
\end{table*}

\subsection{\label{sec:dipole_ME}Dipole matrix elements}

\begin{figure}[htbp]
  \includegraphics[width=0.49\textwidth]{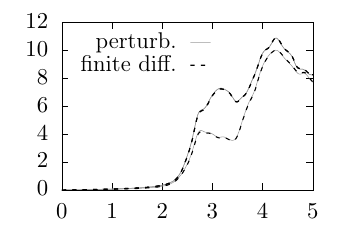}
	\caption{\label{fig:WAVEDER_perturb_vs_LPEAD} Imaginary part of the dielectric permittivity 
	$\varepsilon_2$ on LSDA+$U$ level for two methods of calculating dipole matrix elements, perturbatively and using finite differences.
  }
\end{figure}

Figure~\ref{fig:WAVEDER_perturb_vs_LPEAD} shows $\varepsilon_2$ (broadened by 0.1~eV) on LSDA+$U$ level ($U$=5.3~eV, using PSP0 and $6\times 6\times 6$ $k$-points) 
obtained with dipole matrix elements from perturbation theory and from finite differences, respectively. 
The differences are negligible.

\subsection{\label{sec:TDA}Tamm-Dancoff approximation}

\begin{figure}[htbp]
  \includegraphics[width=0.49\textwidth]{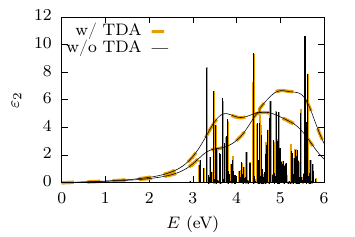}
	\caption{\label{fig:eps2_w_wo_TDA}  Imaginary part of the dielectric permittivity 
	$\varepsilon_2$ with and without the TDA to the BSE.
  }
\end{figure}

Figure~\ref{fig:eps2_w_wo_TDA} shows $\varepsilon_2$ (broadened by 0.4~eV) obtained at the BSE@$G_0W_0$ level with and without the Tamm-Dancoff approximation (TDA),
using $U$=5.3~eV, PSP0, $N_{\omega}$=75, $E_{\mathrm{cut}}(\varepsilon)=266.7$~eV, and 8 valence and 8 conduction bands in the BSE.
The differences due to the TDA are negligible.

\subsection*{Summary}
Approximately converged exciton binding energies can be obtained with the calculation parameters in Table~\ref{tab:converged_calc_param}. 

\begin{table}[htb]
	\begin{tabular}{c|c}
		parameter & converged at \\ \hline
        	$E_{\mathrm{cut}}$   & $\geq$500 eV                                            \\
		$GW$ $k$-mesh  & $4\times 4\times 4$ points (better $6\times 6\times 6$) \\ 
		BSE $k$-mesh  & $6\times 6\times 6$ points (better $10\times 10\times 10$) \\ 
		$N_{\omega}$   & $\geq$100 (better $\geq$ 150)                           \\
		total \# of bands in $GW$ & 1536 (better 2048)                                 \\
        	$E_{\mathrm{cut}}(\varepsilon)$   & $\geq$200 eV                                            \\
		\# of BSE bands & 8 VB, 8 CB (better 12 VB, 12 CB)                        \\
		SOC       & included                                                \\
		local field effects      & included                                                \\
       antiresonant terms & TDA    
	\end{tabular}
	\caption{\label{tab:converged_calc_param}Calculation parameters needed for convergence}
\end{table}

The (BSE@)$G_0W_0$ gaps in the main text were corrected for spin-orbit coupling and extrapolated to an infinite cutoff energy of the dielectric matrix, 
and to an infinitely dense $k$-point mesh (only BSE).


\section{Tests for domain walls}
This section contains convergence tests for the 120-atom cell of BiFeO$_3$ with 71\textdegree~domain walls. 

\subsection{Excitons vs. independent particles in PL spectra}
\begin{figure}[htbp]
	\includegraphics[width=0.49\textwidth]{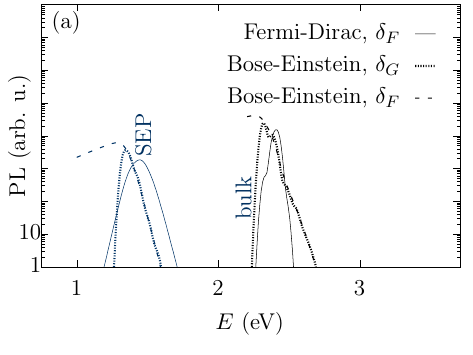}
	\caption{\label{fig:PL_X_vs_IPA}PL spectrum assuming either recombination from excitons (Bose-Einstein distribution) or from independent electrons and holes (Fermi-Dirac distribution). 
	Different approximations of the $\delta$ function in Eq.~(1) in the main text also lead to different shapes ($\delta_F$, $\delta_G$, see text).}
\end{figure}

While the absorption spectra of domain walls were calculated in the independent-particle approximation, 
in the photoluminescence spectrum excitons were introduced afterwards by weighting the absorption spectrum with a Bose-Einstein distribution. 
Figure~\ref{fig:PL_X_vs_IPA} shows the photoluminescence spectrum from this approach in comparison with that from a thermal distribution of independent electrons and holes.
Differences in the spectra arise due to the difference between Bose and Fermi distribution and different approximations of the $\delta$ function in Eq.~(1) in the main text: 
A Gaussian, $\delta(x)\approx\delta_G(x)= A\mathrm{e}^{-x^2/\sigma}$, was tested as well as a Fermi function derivative, $\delta(x)\approx\delta_F(x)= -B df(x)/dx$, $f(x)=1/(\mathrm{e}^{x/k_BT}+1)$.

%
\end{document}